\documentclass[notitlepage,10pt,aps,prd,twocolumn,amsmath,amssymb,floatfix,superscriptaddress,nofootinbib,preprintnumbers]{revtex4-2}

\usepackage{amsfonts,amssymb,amsmath,mathtools,color,bm,epsfig}
\usepackage[normalem]{ulem}
\usepackage{multirow}
\usepackage[section]{placeins}

\definecolor{ultramarine}{rgb}{0.07, 0.04, 0.56}
\definecolor{cadmiumgreen}{rgb}{0.0, 0.42, 0.24}
\definecolor{indigo(dye)}{rgb}{0.0, 0.25, 0.42}
\usepackage[linktocpage=true]{hyperref}
\hypersetup{
colorlinks=true,
citecolor=ultramarine,
linkcolor=cadmiumgreen,
urlcolor=indigo(dye),
}

\renewcommand{\O}{\mathcal{O}}
\newcommand{\ax}{{\rm ax}}
\newcommand{\CDM}{{\rm CDM}}
\newcommand{\eV}{{\rm eV}}
\newcommand{\ef}{{\rm ef}}
\newcommand{\efa}{{\rm efa}}

\def\aeq{a_{\rm eq}}
\def\Heq{H_{\rm eq}}

\newcommand\numberthis{\addtocounter{equation}{1}\tag{\theequation}}
\definecolor{darkgreen}{cmyk}{0.85,0.2,1.00,0.2} 
\definecolor{purple}{cmyk}{0.5,1.0,0,0} 

\begin{document}

\title[]{Accurate effective fluid approximation for ultralight axions}

\author{Samuel Passaglia}
\email{samuel.passaglia@ipmu.jp}
\affiliation{Kavli Institute for the Physics and Mathematics of the Universe (WPI), The University of Tokyo Institutes for Advanced Study (UTIAS),The University of Tokyo, Chiba 277-8583, Japan}

\author{Wayne Hu}
\affiliation{Kavli Institute for Cosmological Physics, Department of Astronomy \& Astrophysics, 
Enrico Fermi Institute, University of Chicago, Chicago, Illinois 60637, USA}

\label{firstpage}

\begin{abstract}
Ultralight axions are theoretically interesting and phenomenologically rich dark sector candidates, but they are difficult to track across cosmological timescales because of their fast oscillations. We resolve this problem by developing a novel method to evolve them efficiently and accurately. We first construct an exact effective fluid which at late times matches the axion but which evolves in a simple way. We then approximate this evolution with a carefully chosen equation of state and sound speed. With our scheme we find that we can obtain subpercent accuracy for the linear theory suppression of axion density fluctuations relative to that of cold dark matter without tracking even a single complete oscillation of the axion field. We use our technique to test other approximation schemes and to provide a fitting formula for the transfer function for the matter power spectrum in linear theory in axion models. Implementing our approach in existing cosmological axion codes is straightforward and will help unleash the potential of high-precision next-generation experiments.
\end{abstract}

\date{\today}

\maketitle 

\section{Introduction}

Axions are hypothetical particles that might reside in the dark sector of $\Lambda$CDM \cite{Marsh:2015xka} and which can arise in many different physical scenarios with a wide range of cosmologically interesting masses \cite{Peccei:1977hh,Svrcek:2006yi,Arvanitaki:2009fg}. For many cosmological purposes axion particles behave simply as a classical scalar field, with the field's initial displacement in the potential and mass determining their cosmological evolution. While the Hubble rate $H$ is larger than the axion mass $m$, the field remains effectively frozen. Once the Hubble rate drops below the mass the field oscillates on a timescale $m^{-1}$, redshifting like cold dark matter (CDM) across Hubble timescales $H^{-1}$.

A critical mass is therefore the Hubble rate at matter-radiation equality $\Heq \sim 10^{-28} \ \eV$. Lighter axions than this behave more like dark energy \cite{Frieman:1995pm,Choi:1999xn,Marsh:2010wq} and heavier ones more like dark matter \cite{Preskill:1982cy,Turner:1983he,Brandenberger:1984jq,Khlopov:1985jw,Ratra:1987rm}. Between $\Lambda$ and CDM-like limits, axions can be responsible for a wide variety of interesting new cosmological signatures beyond $\Lambda$CDM (see Refs.~\cite{Grin:2019mub,Ferreira:2020fam} and references therein), for example acting as early dark energy components \cite{Poulin:2018dzj,Lin:2019qug} or sourcing isocurvature perturbations in the CMB \cite{Fox:2004kb}. 

One of the most important signatures of ultralight axions is that they suppress small-scale gravitational clustering due to their macroscopic de Broglie wavelengths, which occurs on the kiloparsec scale for $m \sim 10^{-22} \ \eV$  \cite{Hu:2000ke}, and leads to a host of observational consequences (see, e.g., Ref.~\cite{Hui:2016ltb} and references therein). However, it is highly challenging to track the oscillations of these axions across a Hubble time today $H_0^{-1} \simeq (10^{-33} \ \eV)^{-1}$ in order to make accurate theoretical predictions.

This difficulty is usually addressed with an effective fluid approximation (EFA)  that replaces the exact Klein-Gordon solution for the field by effectively averaging over the axion oscillations \cite{Ratra:1990me,Hu:2000ke,Hlozek:2016lzm,Farren:2021jcd}. This approximation inevitably induces errors when making predictions for observables because the axion field is initially effectively frozen and the approximate equations of state used to evolve the effective fluid efficiently may then be insufficiently accurate. Similar difficulties arise if the Klein-Gordon equation is instead recast as a Schr{\"o}dinger equation \cite{Hsu:2020ikn,Zhang:2017flu,Zhang:2017dpp,Salehian:2020bon,Namjoo:2017nia}.

Even for axions in the range $10^{-27} \ \eV \lesssim m \lesssim 10^{-24} \ \eV$ where the clustering suppression in linear theory is testable with CMB measurements (see, e.g., Refs.~\cite{Amendola:2005ad,Hlozek:2017zzf}) and the timescale hierarchy is less extreme, these errors have already been shown to be significant at the $1\sim2 \sigma$ level \cite{Cookmeyer:2019rna} for upcoming experiments like CMB-S4 \cite{Abazajian:2019eic}. Other similar averaging approaches such as the one proposed by Ref.~\cite{Urena-Lopez:2015gur} have been shown to suffer the same problems \cite{Cookmeyer:2019rna}. 

For heavier axions $10^{-24} \ \eV \lesssim m \lesssim 10^{-19} \ \eV$, where the hierarchy is larger, the error has not yet even been characterized. On the low-mass end these axions are best probed by their effect on large-scale structure in existing \cite{Dentler:2021zij} and next-generation optical and infrared survey experiments \cite{Aghamousa:2016zmz,Amendola:2016saw,Dore:2018smn}. On the high-mass end they are subject to subgalactic tests \cite{Chen:2016unw,Hayashi:2021xxu} and to early-time structure probes like 21cm \cite{Bozek:2014uqa,Lidz:2018fqo,Jones:2021mrs,Sarkar:2022dvl} and the Lyman-$\alpha$ forest \cite{Irsic:2017yje}. In this regime linear theory results are necessary as initial conditions for nonlinear simulations (see, e.g., Refs.~\cite{Schive:2014hza,Veltmaat:2018dfz,Mocz:2017wlg}).

In this work, we develop a method to dramatically reduce the error made by the EFA in linear theory by both judiciously choosing how the effective fluid (EF) is constructed from the exact field solution, and by improving the approximate equation of state and sound speed used to evolve the effective fluid.  In \S\ref{sec:method}, we present and validate our method at both the background \S\ref{subsec:background} and perturbation \S\ref{subsec:perturbations} levels. In \S\ref{sec:results}, we implement our approach to compute the matter power spectrum transfer function in axion models and to quantify the error made by a usual fluid approximation. We compare to the approach taken by \textsc{AxionCAMB} in Appendix~\ref{app:axioncamb}. We conclude in \S\ref{sec:conclusion}. 

\section{Method}
\label{sec:method}

The axion field $\phi$ obeys the Klein-Gordon equation,
\begin{equation}
\Box \phi(\vec{x}, t) = V'(\phi(\vec{x}, t) ) \approx m^2 \phi(\vec{x}, t)\,,
\label{eq:KG}
\end{equation}
where the approximation of the axion cosine potential with a quadratic potential $V \approx m^2\phi^2/2$ applies sufficiently near the minimum.

Given that this is a wave equation with solutions that oscillate on the mass timescale $t \sim m^{-1}$, we wish to find a computationally tractable way to solve it across cosmological timescales $H_0^{-1}$ to a given accuracy. We restrict ourselves to linear theory, splitting the axion field $\phi(\vec{x}, t)$ into a background $\phi(t)$ and linear perturbations $\delta \phi(\vec{x}, t)$, and we develop a procedure for each piece separately.

Schematically, our approach is to decompose the field $\phi$ into two auxiliary fields $\varphi_{c, s}$ which factor out the mass timescale oscillations starting at some switch time in the oscillatory regime which we mark with a subscript `*' and which we parametrize by the ratio $m/H_* > 1$. 

For the background, we use a specific combination of these auxiliary field variables to construct an effective fluid with density $\rho_\ax^\ef$ in such a way that at late times it approaches the true energy density of the axion $\rho_\ax$ up to a small matching error suppressed by $(m/H_*)^{-3}$.

The effective fluid can be evolved exactly, but it has the advantage that it does not significantly evolve on the mass timescale. Therefore its evolution from early to late times can be accurately approximated in a straightforward and computationally efficient manner with an effective fluid approximation, yielding quantities such as the effective fluid approximation density $\rho_\ax^{\efa}$ which approximates its exact counterpart $\rho_\ax^\ef$ up to a small evolution error which is also suppressed by $(m/H_*)^{-3}$. 

We are therefore able to approximate the late-time axion fluid by evolving efficiently only our effective fluid approximations after $m/H_*$. We can control the matching and evolution errors to any desired accuracy by choosing the switch epoch $m/H_*$. We develop a similar procedure for the perturbations, where there are additional terms associated with Jeans oscillations, and we find that a very modest switch parameter value $m/H_* = 10$ is already sufficient to obtain subpercent accuracy for observable quantities like the axion transfer function out to scales where the linear power remains appreciable. The parameter value $m/H_* = 10$ corresponds to solving less than one oscillation of the axion field exactly, after which the axion is evolved only in the effective fluid approximation.

\subsection{Background}
\label{subsec:background}

The equation of motion for the background is
\begin{equation}
\label{eq:bgKG}
\ddot\phi + 2 
\frac{\dot a}{a}
\dot\phi + a^2 m^2 \phi = 0\,,
\end{equation}
with overdots denoting derivatives with respect to the conformal time $\eta =\int dt/a$. At early times $m/H \ll 1$ the field is frozen by Hubble drag and behaves like a dark energy component to the universe's energy budget. As $H$ drops below $m$, the field is released and oscillates around its potential minimum, with its energy density redshifting as cold dark matter once $m/H \gg 1$.

When the expansion rate is a power law in time, the background \eqref{eq:bgKG} has an exact solution in terms of Bessel functions (see, e.g., Ref.~\cite{Marsh:2010wq}). We can employ this solution in the radiation dominated regime, but in order to also establish some groundwork for the analysis of perturbations, which do not admit such a solution, we instead focus on the WKB solution for the background in the oscillatory regime. The squared frequency of the oscillator $a^2 m^2 - \ddot{a}/a$ goes to $a^2 m^2$ in radiation domination or when $m/H\gg1$, so the asymptotic solution is a harmonic oscillator
\begin{equation}
\phi \rightarrow \frac{C_1}{m^{1/2}a^{3/2}} \cos \left[ \tau + C_2 \right]\,,
\end{equation}
where $\tau \equiv mt$ is a convenient time variable in the oscillatory regime and $C_1$ and $C_2$ are constants determined by the initial conditions.

The oscillating field admits a
description in terms of its energy density and pressure,
\begin{align*}
\label{eq:rhoP}
\rho_\ax &\equiv \frac{1}{2} \left(\frac{d \phi}{d t}\right)^2  + V\,, \\
P_\ax &\equiv \frac{1}{2} \left(\frac{d \phi}{d t}\right)^2  - V\,,
\numberthis
\end{align*}
which approach
\begin{align*}
\label{eq:cdm}
a^3 \rho_\ax &\propto  1 +\O\left(  \cos\left[2 \tau\right]  \left(\frac{m}{H}\right)^{-1},\left( \frac{m}{H}\right)^{-2} \right)\,, \numberthis\\
a^3 P_\ax &\propto  \cos\left[2 {(\tau + C_2)}\right] +\O\left(  \cos\left[2 \tau\right]   \left(\frac{m}{H}\right)^{-1},\left(\frac{m}{H}\right)^{-2}\right)\,
\end{align*}
where the $\cos[2 \tau]$ factors here signify that the neglected terms oscillate on the $2 \tau$ timescale; they are not meant to imply a specific phase of the oscillations in the neglected term. 

On the cycle average, this representation of the axion looks like pressureless matter to order $\O(m/H)^{-2}$, so the usual approach to avoiding the axion's timescale hierarchy used in state-of-the-art codes like \textsc{AxionCAMB} \cite{Hlozek:2014lca} is to solve the Klein-Gordon equation exactly until the oscillations begin, with the default setting in \textsc{AxionCAMB} defining this as $m/H=3$, and then compute the axion energy density and evolve it forward as a pressureless fluid. 

However, this procedure inevitably chooses an arbitrary phase in the axion oscillation at the matching point between the field and the fluid. Eq.~\eqref{eq:cdm} shows that these oscillatory modes are large, introducing an error $\propto (m/H)^{-1}$ at the matching time. This leads to a matching error in the axion density that remains today.

Moreover, accurately evolving the axion density requires knowing the exact form of the $\O(m/H)^{-2}$ term in Eq.~\eqref{eq:cdm} in order to give the fluid the appropriate leading-order correction to the CDM-like equation of state. While at the background level the appropriate equation of state is straightforward to derive analytically, at the perturbation level the situation is more complicated and we would like to have an empirical method of calibrating the appropriate fluid evolution.

In this work, we construct an effective fluid approximation in such a way as to control both the matching errors and the evolution errors. To do so, we start from some time $\tau_*$ in the oscillatory regime, corresponding to the ratio $m/H_* > 1$, and write the field $\phi$ in a form which factors out subsequent oscillations,
\begin{equation}
\label{eq:ansatz}
\phi(\tau) = \varphi_c(\tau) \cos\left[\tau -\tau_* \right] + \varphi_s(\tau) \sin\left[\tau - \tau_* \right]\,,
\end{equation}
by using two auxiliary field variables $\varphi_{c,s}$. This decomposition is useful because if $\varphi_{c,s}$ evolve only on the Hubble timescale, then they can be used to compute quantities which should look like cycle averaged versions of the axion. 

To see how $\varphi_{c,s}$ evolve, we plug the ansatz \eqref{eq:ansatz} back into the equation of motion \eqref{eq:bgKG} and obtain an equation of motion for $\varphi_{c,s}$, 
\begin{align*}
\label{eq:bgKGvar}
\left(\varphi_c'' + 2 \varphi_s' + 3 \frac{H}{m} [\varphi_s + \varphi_c' ]\right) \cos\left[\tau -\tau_* \right] \ &+\\ \left( \varphi_s''- 2 \varphi_c' + 3 \frac{H}{m} [-\varphi_c + \varphi_s']\right)  \sin \left[\tau - \tau_* \right] &= 0\,,
\numberthis
\end{align*}
which we can impose is solved by setting the two individual terms in the parentheses to zero. Primes $'$ here and throughout denote derivatives with respect to $\tau$. As long as the initial conditions for $\varphi_{c,s}$ and their derivatives at $\tau_*$ are chosen appropriately so that they match $\phi(\tau_*)$ and $\phi'(\tau_*)$ through Eq.~\eqref{eq:ansatz}, then the $\varphi_{c,s}$ which solve this auxiliary equation of motion are exact representations of the axion background.

\begin{figure}[t]
\includegraphics[]{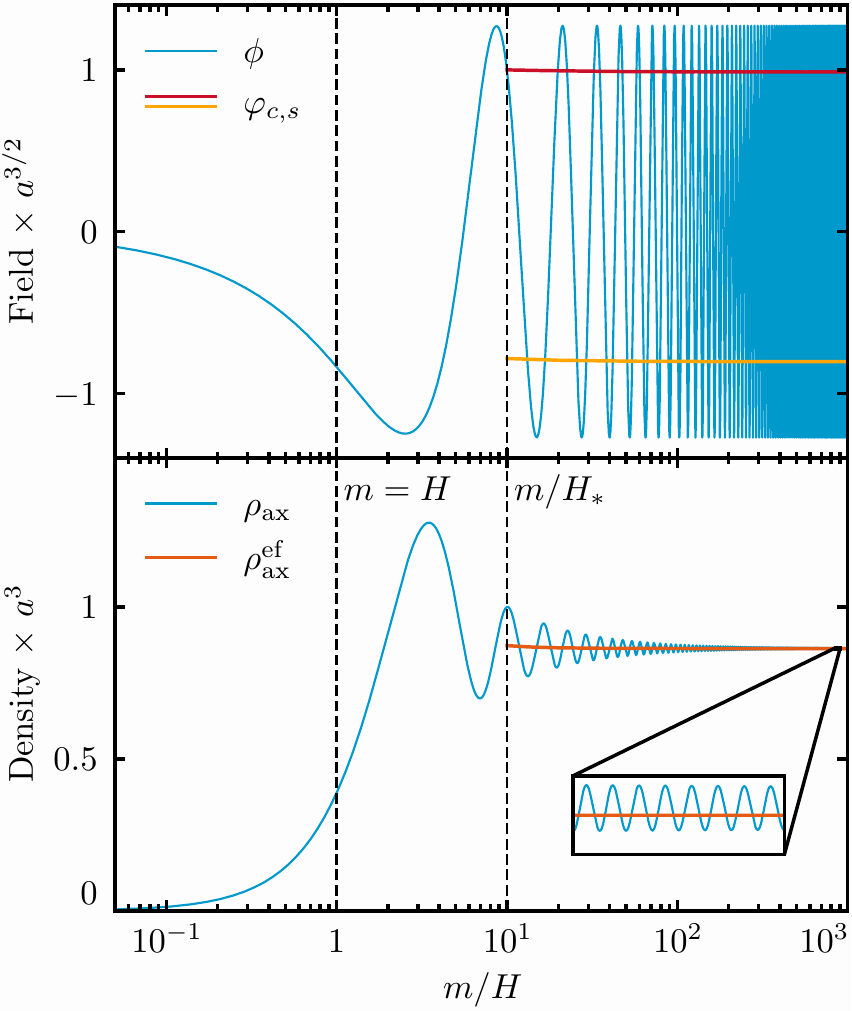}
\caption{The axion field $\phi$ (upper panel, blue) is released from Hubble drag around $m\sim H$ and then oscillates around its minimum as $m/H$ increases. At some arbitrary $m/H_*$ in the oscillatory phase, we write $\phi$ in terms of auxiliary fields $\varphi_{c,s}$ (red and yellow curves) which factor out oscillations. These auxiliary fields allow us to construct the effective fluid density $\rho_\ax^\ef$ (lower panel, orange), which matches the axion density $\rho_\ax$ (blue) at late times up to a tiny matching error (inset), which our analytic results show is imprinted at the switch time and is proportional to $(m/H_*)^{-3}$. The evolution of $\rho_\ax^\ef a^3$ is very small and therefore its value at late times can be accurately computed from its early time value
using an approximate equation of state. Variables here are scaled by the field and its density at $m/H_*$ and their asymptotic behavior in the $m/H \rightarrow \infty$ limit, the axion mass is $10^{-22}\ \eV$, and its present-day abundance is that of the dark matter. For further discussion see \S\ref{subsubsec:matching}.}
\label{fig:background}
\end{figure}

We now use our $\varphi_{c, s}$ variables to {\it define} effective fluid quantities
\begin{align*}
\label{eq:rhoPef}
\rho_\ax^\ef &\equiv  \frac{1}{2} m^2 \left(\varphi_c^2 +  \varphi_s^2  + \frac{{\varphi_c'}^2}{2} +\frac{{\varphi_s'}^2}{2}  -  \varphi_c \varphi_s' +  \varphi_s \varphi_c'\right)\,,  \\
P_\ax^\ef &\equiv \frac{1}{2} m^2 \left(\frac{{\varphi_c'}^2}{2} +\frac{{\varphi_s'}^2}{2}  -  \varphi_c \varphi_s' +  \varphi_s \varphi_c'\right)\,, \numberthis
\end{align*}
which we have constructed using the Klein-Gordon equation \eqref{eq:bgKGvar} to exactly satisfy the usual form of a fluid conservation law (e.g.~Ref.~\cite{Hu:1998kj})
\begin{equation}
\label{eq:ef_bg}
\dot\rho_\ax^\ef = -3 {\frac{\dot{a}}{a}}\left( \rho_\ax^\ef   +P_\ax^\ef \right)\,.
\end{equation}
The definitions \eqref{eq:rhoPef} of the effective fluid are motivated by the notion of cycle averaging the oscillations in $\tau$ and they can be derived by plugging the ansatz for $\phi$ \eqref{eq:ansatz} into the energy density and pressure equations \eqref{eq:rhoP} and then sending $\sin^2$ and $\cos^2$ terms to $1/2$ and cross terms to zero. Regardless of their approximate motivation, however, the evolution equation \eqref{eq:ef_bg} they yield for the effective fluid density $\rho_\ax^\ef$ is exact. To evaluate it and close the system requires knowing the evolution of $P_\ax^\ef$ through $\varphi_{c, s}$ using the Klein-Gordon equation.

So far we have simply recast the axion system using different variables with no approximations, but doing so has two advantages.  First, at late times the effective fluid density is an excellent approximation for the true axion density. We call a matching error any late-time difference between the two densities, and we will show this error can be efficiently controlled. Second, $\rho_\ax^\ef$ does not have significant oscillations as it evolves and therefore its value at late times can be easily approximated by evolving its value at early times using an equation of state $w_\ax^\efa$ which approximates $P_\ax^\ef/\rho_\ax^\ef$. We label such an approximate effective fluid $\rho_\ax^\efa$, and we call any difference between the effective fluid $\rho_\ax^\ef$ and the effective fluid approximation $\rho_\ax^\efa$ an evolution error. We will show that the evolution error is suppressed once $w_\ax^\efa$ is appropriately calibrated. To the extent that both the matching and evolution errors can be made negligible, then we no longer need to solve the Klein-Gordon equation after $\tau_*$ and the technique as a whole becomes extremely efficient and accurate.

We now quantify the matching and evolution errors in turn.

\subsubsection{Matching error}
\label{subsubsec:matching}

We first show in Fig.~\ref{fig:background} how a matching error appears in practice with numerical solutions of the exact equations of motion. In the Hubble drag regime and for the first few oscillations, we solve the Klein-Gordon equation \eqref{eq:bgKG} for the axion field $\phi$ directly. At $m/H_*$, chosen here to be $10$, we switch to solving for the auxiliary field variables $\varphi_{c, s}$ using the two pieces of Eq.~\eqref{eq:bgKGvar}. These factor out the $\phi$ oscillations (upper panel), and enable us to compute our effective fluid density $\rho_\ax^\ef$ at subsequent times exactly (lower panel). For clarity, we have scaled the field values and densities with $a^{3/2}$ and $a^3$ to compensate their asymptotic behavior and normalized the result at $m/H=10$.

The effective fluid density $\rho_\ax^\ef$ is an exact quantity in the sense that it is constructed from the exact axion field solution of the equation of motion, but at late times it does not represent exactly the true axion density $\rho_\ax$, as shown by the small difference between $\rho_\ax^\ef$ and the cycle-average of $\rho_\ax$ in the lower panel inset. This difference does not go away as $m/H \rightarrow \infty$.  It is the matching error made by our scheme, which we will show is imprinted at the switch time and suppressed by $\O(m_*/H)^{-3}$. We can therefore make it arbitrarily small by setting the switch time sufficiently late.

We quantify the matching error by studying an exact power-series solution to the Klein-Gordon equation. When the axion is a negligible component of the energy budget and the background is externally determined as some $a(\tau) \propto \tau^p$, the auxiliary equation of motion \eqref{eq:bgKGvar} admits solutions of the form 
\begin{align*}
\label{eq:bgsol}
 a^{3/2} \varphi_c(\tau) &=  C(\tau) + \begin{pmatrix}A(\tau), & B(\tau)\end{pmatrix} 
 \begin{pmatrix} 
 \cos\left[2(\tau - \tau_*)\right] \\ 
 \sin\left[2(\tau - \tau_*)\right]
\end{pmatrix}\,,   \\
\numberthis 
\\
  a^{3/2} \varphi_s(\tau) &=  S(\tau)  + \begin{pmatrix}-B(\tau), &A(\tau)\end{pmatrix} 
 \begin{pmatrix} 
 \cos\left[2(\tau - \tau_*)\right] \\ 
 \sin\left[2(\tau - \tau_*)\right]
\end{pmatrix}\,,
\end{align*}
where $A$, $B$, $C$, and $S$ here are coefficients which evolve only on the Hubble timescale, satisfying power series solutions in $\left(m/H\right)^{-n}$ and approaching constants at late times $m/H \rightarrow\infty$. The coefficients $A$ and $B$ represent fast oscillatory modes of $\varphi_{c, s}$ and hence naively violate our expectations that $\varphi_{c,s}$ evolve only slowly, but really they reflect a redundancy in our decomposition \eqref{eq:ansatz} due to the trigonometric identities
\begin{align*}
\label{eq:trig}
\cos(x) ={}& \cos(2 x) \cos(x) + \sin(2 x) \sin(x)\,, \\
\sin(x) ={}& \sin(2 x) \cos(x) - \cos(2 x) \sin(x)\,, \numberthis
\end{align*}
which allows a remapping which sets them to zero
\begin{align*}
C+A \rightarrow{}& C , \quad A\rightarrow 0\,, \\
S+B \rightarrow{}& S, \quad B\rightarrow 0\,.\numberthis
\label{eq:ABdegeneracy}
\end{align*}

\begin{figure}[t]
\includegraphics[]{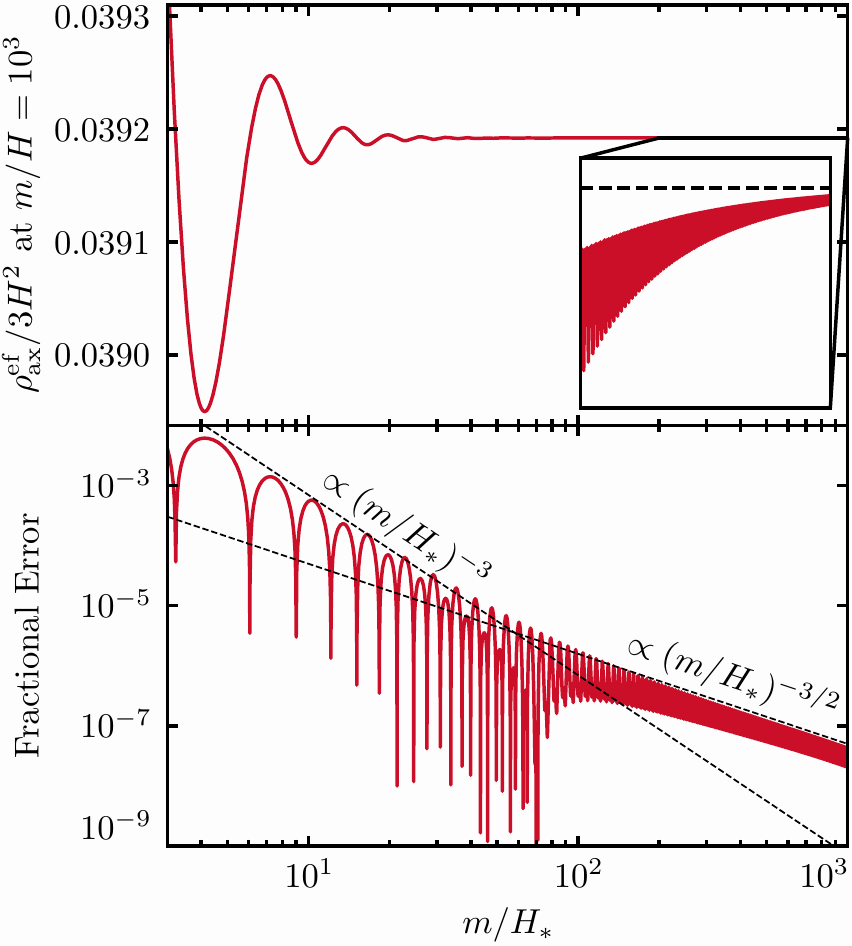}
\caption{The effective fluid density $\rho_\ax^\ef$ evaluated at a fixed time, here chosen to be $m/H = 10^3$, converges as the switch time parameterized by $m/H_*$ becomes larger (upper panel). Relative to the estimated asymptotic value (inset horizontal, dashed line), this matching error
is already smaller than $0.1\%$ for an early switch $m/H_*= 10$, with the error subsequently decreasing as $(m/H_*)^{-3}$ as predicted by our analytics for a tracer field (lower panel). At the level where the backreaction of the axion on Hubble becomes important, the improvement slows to $(m/H_*)^{-3/2}$. For further discussion see \S\ref{subsubsec:matching}.} 
\label{fig:background_accuracy}
\end{figure}

Given a generic matching condition at $\tau_*$, however,
these oscillations in the auxiliary variables produce a constant\footnote{Even though the terms inherited from $A$ and $B$ oscillate with a constant fractional correction to the field (Eq.~\eqref{eq:bgsol}) as $m/H\rightarrow \infty$, these oscillations are suppressed by an additional $(m/H)^{-1}$ term in the fluid quantities Eq.~\eqref{eq:rhoPef} due to the combination of time derivative terms.}  fractional offset between the effective fluid density \eqref{eq:rhoPef} and the true density \eqref{eq:rhoP} at late times once they are imprinted at $\tau_*$. This occurs because given the remapping of Eq.~(\ref{eq:ABdegeneracy}), $C+A$ and $S+B$ should add coherently in the squared time average for $\rho_{\ax}$ whereas Eq.~(\ref{eq:rhoPef}) for $\rho_\ax^\ef$ drops the cross terms. Therefore the late time $\rho_\ax^\ef$ will disagree with the late time $\rho_\ax$  due to contributions from $AC$ and $BS$. This is a matching error.
 
We therefore want to make the $A$ and $B$ modes as small as possible. We do so by choosing an appropriate matching condition at the switch time $\tau_*$. When we switch from the $\phi$ variable to the auxiliary $\varphi_{c, s}$ variables, we must provide two arbitrary additional constraints beyond the two exact matching conditions for $\phi$ and $\phi'$ to determine the four initial conditions for $\varphi_{c, s}$ and $\varphi_{c, s}'$. We can choose these such that $A$ and $B$ are suppressed. Given that we know the $\varphi_{c,s} \propto a^{-3/2}$ evolutionary form of the $C$ and $S$ terms, we impose initial constraints of the form\footnote{
$\langle H \rangle$ here denotes the Hubble rate averaged over axion oscillation cycles, a distinction only relevant when the axion is energetically relevant at $\tau_*$, which is beyond the limits of our analytic analysis. 
}
\begin{align*}
\label{eq:constraints}
\left.\frac{\varphi_{c,s}''}{\varphi_{c,s}'}\right\vert_{*} = \left.-\frac{1}{2} \frac{\langle H \rangle}{m} \left(3 - \frac{m}{\langle H\rangle ^3}\frac{d\langle H\rangle^2}{d \tau}\right)\right\vert_{*} \,, \numberthis
\end{align*}
which suppresses the oscillatory modes as
\begin{align*}
\label{eq:ABsuppression}
\frac{A}{S} \sim \frac{B}{C} \sim  \O\left(\frac{m}{H_*}\right)^{-3}\,. \numberthis
\end{align*}
Initial conditions that yield an even larger suppression of the oscillatory modes can also be constructed\footnote{At next order, imposing
\begin{align*}
\left. \varphi_{c,s}''\right\vert_{*} &= \left(-\frac{1}{2} \frac{\langle H \rangle}{m} \left(3 - \frac{m}{\langle H\rangle ^3}\frac{d\langle H\rangle^2}{d \tau}\right)\right) \times \\
&\left. \left(\varphi_{c,s}' + \left(\varphi_{c,s}'  + \frac{3 \langle H \rangle}{2 m} \varphi_{c,s}\right)\right) \right\vert_{*} \,, 
\end{align*}
yields a suppression of order $\O\left(m/H_*{}\right)^{-4}$.}, but for our purposes we will see that $\O\left(m/H_*{}\right)^{-3}$ is sufficient.

With this suppression of the fast oscillatory modes of $\varphi_{c,s}$, the relationship between our effective fluid variable and the true fluid variables at late times is
\begin{equation}
\label{eq:matching_error_BG}
\rho_\ax^\ef = \rho_\ax \times \left(1 + \mathcal{O}\left( \frac{m}{H_*}\right)^{-3} \right) \quad (m/H \gg 1)\,.
\end{equation}
This is the expected amplitude of the discrepancy between the true and effective fluids at late times that we saw in Fig.~\ref{fig:background}. There we chose a switch time $m/H_* = 10$ and the discrepancy is roughly $\sim 10^{-3}$. This matching error can then be reduced by making $m/H_*$ larger and the switch time later.

We confirm numerically in Fig.~\ref{fig:background_accuracy} the analytically derived $(m/H_*)^{-3}$ scaling of our matching error \eqref{eq:matching_error_BG}. Since the matching error must vanish as the switch time $m/H_* \rightarrow \infty$, we can estimate it by checking how $\rho_\ax^\ef$ evaluated at a fixed late time depends on the switch time $m/H_*$. This is the upper panel, where we show the effective fluid density $\rho_\ax^\ef$ evaluated at a time corresponding to $m/H = 10^3$ as a function of the switch epoch $m/H_*$. As $m/H_*$ is taken deeper in the oscillatory regime, $\rho_\ax^\ef$ converges to a value which we estimate with the horizontal line --- the exact value is not important. The distance from this line for a given $m/H_*$ is then the matching error which we show in the lower panel fractionally. Phrased in terms of Fig.~\ref{fig:background}, Fig.~\ref{fig:background_accuracy} shows how the orange line in the inset changes as a function of the switch epoch $m/H_*$. 

Very modest switch epochs $m/H_* \sim 10$ result in a matching error of less than $ 0.1\% $. The matching error then improves at first as the $(m/H_*)^{-3}$ power law we derived in Eq.~\eqref{eq:matching_error_BG}. As $m/H_*$ increases, the modulation of the error amplitude between positive and negative oscillation extrema indicates a drifting mean and eventually the error is dominated by the secular drift rather than an oscillatory error. In this regime the accuracy is already excellent but the error decreases at a slower rate $\propto(m/H_*)^{-3/2}$. We have confirmed numerically that this change in convergence rate is due to the backreaction of the axion on its own evolution through the Hubble rate, an effect which we did not include when deriving our analytic scalings. 

This slower scaling with $m/H_*$ begins when the matching error is already extremely small. For a fixed present-day abundance, axions which are lighter than the $10^{-22} \ \eV$ axion shown here will be more energetically important at a given $m/H$ and this convergence slowdown would therefore become important earlier. However, CMB constraints for masses where $m/H \sim 1$ at matter-radiation equality are already powerful enough that such axions are limited to just a small component of the dark matter \cite{Hlozek:2014lca}. Consequently their effect on the Hubble rate and therefore this slowdown in our accuracy improvement is suppressed. Similarly small scale structure constraints such as the Lyman-$\alpha$ forest also limit the fraction of dark matter for more intermediate-mass axions \cite{Irsic:2017yje}. 

We have now constructed an effective fluid which at late times matches the true fluid to a very good accuracy. So far we have been evolving this effective fluid exactly. We now approximate its evolution and study the resulting evolution error.

\subsubsection{Evolution error}
\label{subsubsec:evolution}

\begin{figure}[t]
\includegraphics[]{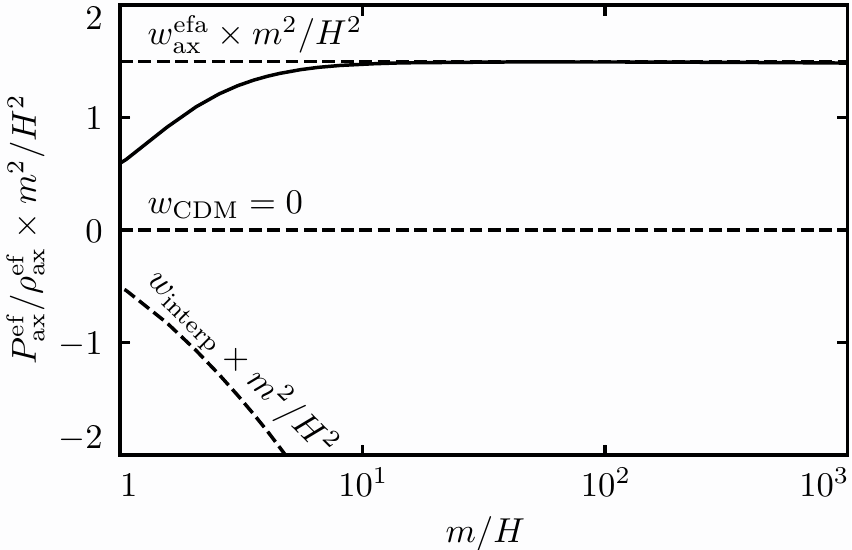}
\caption{The exact equation of state of the effective fluid $P_\ax^\ef / \rho_\ax^\ef$ (solid) with an $(m/H)^{-2}$ scaling factored out. We choose to approximate it with $w_\ax^\efa = 3/2 (m/H)^{-2}$, which is the asymptotic behavior of $P_\ax^\ef / \rho_\ax^\ef$ in radiation domination as can also be shown analytically. Other choices such as the pressureless equation of state $w_{\CDM} = 0$ and especially the interpolating equation of state $w_{\rm interp}$ \eqref{eq:wax_interpolate} are suboptimal. The difference at late times between $w_\ax^\efa$ and $P_\ax^\ef / \rho_\ax^\ef$ is due to the transition to matter domination, but is insignificant due to the $(m/H)^{-2}$ suppression. The effective fluid here is constructed at the time $m/H_* = m/H$ for each point on the solid curve. For further discussion, see \S\ref{subsubsec:evolution}.}
\label{fig:w}
\end{figure}

Evolving $\rho_\ax^\ef$ exactly requires knowing $P_\ax^\ef$ exactly through the auxiliary variables $\varphi_{c,s}$. These are not easier to evolve than the usual axion field $\phi$ due to the small $2\tau$ oscillations from the matching error.  We therefore want to approximate the effective fluid conservation law \eqref{eq:ef_bg} by replacing the exact $P_\ax^\ef/\rho_\ax^\ef$ with an approximate equation of state $w_\ax^\efa$ at the switch epoch defined by the desired $m/H_*$.

The simplest equation of state to use would be the pressureless value $w_{\CDM} = 0$. This is the fluid evolution used in \textsc{AxionCAMB}, and from the leading order WKB solution for the field, Eq.~\eqref{eq:cdm}, we expect $w_{\CDM}$ to make an instantaneous error of order $(m/H)^{-2}$ which will then be integrated from the switch epoch $(m/H_*)$ to today. Since this is larger than our matching error $\O(m/H_*)^{-3}$, we wish to find a better equation of state.

\begin{figure*}[t]
\includegraphics[]{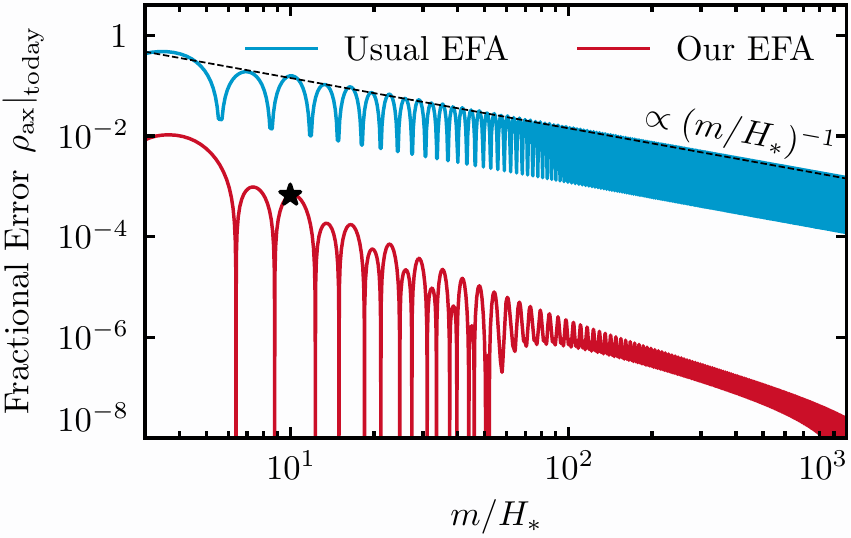}
\includegraphics[]{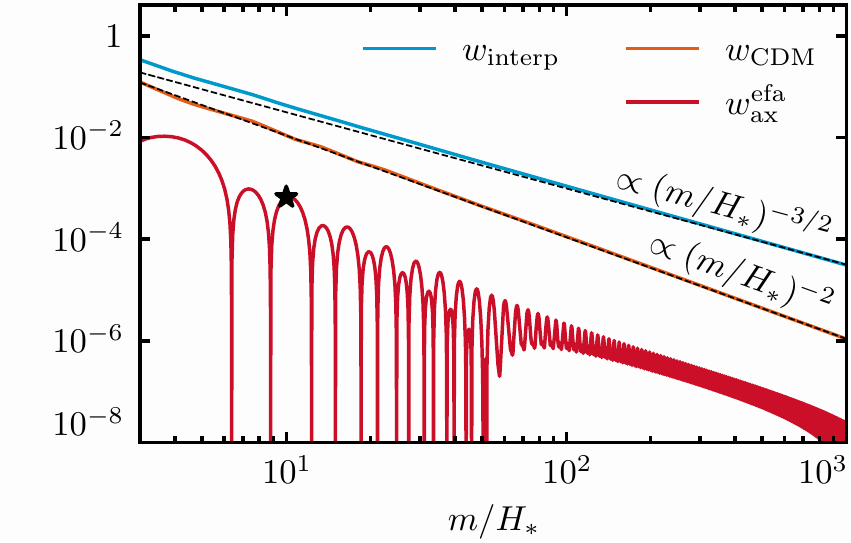}
\caption{Left: Our effective fluid approximation (red) dramatically decreases the error in the axion density today $\rho_{\ax}\vert_{\rm today}$ as compared to the usual effective fluid approximation (blue) often constructed in the literature by matching the instantaneous energy density of the axion at the switch epoch $m/H_*$. The star denotes our reference solution scheme in which we switch from solving the Klein-Gordon equation to the EFA at $m/H_* = 10$ and therefore achieve subpercent accuracy at the level of the background. Our approach also shows an accelerated convergence rate relative to the usual EFA. Right: The convergence rate of our approach depends on the approximate equation of state we use to evolve the effective fluid, and although $w_{\CDM}=0$ (orange) performs better than an interpolating $w_{\rm interp}$ \eqref{eq:wax_interpolate} (blue) for switches in the oscillatory regime, our choice $w_{\ax}^\efa$ \eqref{eq:wax_next} (red) is so good that the matching error from Fig.~\ref{fig:background_accuracy} dominates over this evolution error. The accuracy in this figure is measured relative to a very late switch time $m/H_* = 10^4$. For further discussion see \S\ref{subsubsec:evolution}.} 
\label{fig:background_fluid_accuracy}
\end{figure*}

One choice often used in the literature is an interpolating form \cite{Poulin:2018dzj}
\begin{equation}
\label{eq:wax_interpolate}
w_{\rm interp} = \frac{1}{1+\left(a_c/a\right)^3}-1\,,
\end{equation}
which approaches $w_{\CDM}$ at late times while acting as a dark energy-like $w_{\rm DE} = -1$ component in the drag regime. $a_c$ is the scale factor when $H=m$. This form is used, for example, in the analysis of Ref.~\cite{Poulin:2018dzj} implemented in the code \textsc{AxiCLASS}. 

In the late time $a_c/a \ll1$ regime the interpolating equation of state approaches $0$ from below as $-(m/H)^{-3/2}$ in radiation domination. However it was not constructed to have the correct approach to $0$, and in fact this approach has not only the wrong exponent but also the wrong sign. It thus makes a very large evolution error and for our purposes is worse than $w_{\CDM}$.

Instead, to find a good choice for the equation of state we can choose to calibrate $w_\ax^\efa$ using the exactly computed $P_\ax^\ef$.

Due to the $m/H_*$ dependent matching error between the effective fluid and the true axion, the exact $P_\ax^\ef$ depends on $m/H_*$. Since we minimized the matching error, however, this effect is insignificant and we can find a sufficiently good approximation for $P_\ax^\ef / \rho_\ax^\ef$ without attempting to fit the $m/H_*$ dependence. In effect we seek to minimize the evolution error in the absence of matching error since the true equation of state for the effective axion fluid should not depend on the matching error.

We therefore plot $P_\ax^\ef/\rho_\ax^\ef$ as a function of $m/H$ while minimizing the matching errors by constructing the effective fluid at $m/H_* = m/H$, i.e.\ by constructing the effective fluid at the time at which we wish to know the equation of state. We show this quantity in Fig.~\ref{fig:w}, along with our choice of approximation
\begin{equation}
\label{eq:wax_next}
w_\ax^\efa = \frac{3}{2} \left(\frac{m}{H}\right)^{-2}\,.
\end{equation}

These show excellent agreement already at $m/H = 10$. The deviation at very late times is due to the eventual transition to matter domination, but for dark-matter like axions this occurs far in the oscillatory regime where the equation of state is suppressed by $(m/H)^{-2}$. 

In fact, the result that the axion equation of state is asymptotically $(3/2) (m/H)^{-2}$ in radiation domination can be shown directly from a second-order WKB solution for $\phi$ (e.g.\ Ref.~\cite{Bender:1999awr}) or from the exact Bessel function solution. Our approach of directly fitting $P_\ax^\ef / \rho_\ax^\ef$ has the advantage that it does not require an analytic solution for the system. This will be useful when we turn our attention to the perturbations, where exact solutions are not available.

Since we now know the appropriate $w_\ax^\efa$ needed to evolve the EFA, our strategy is now to choose $m/H_*$ large enough to reduce the matching error $\O(m/H_*)^{-3}$ to a desired level, compute $\varphi_{c,s}$ and the effective fluid density $\rho_\ax^\ef$ at that time, and then evolve it forward using the effective fluid conservation equation \eqref{eq:ef_bg} with our approximate equation of state $w_\ax^\efa$. This defines $\rho_\ax^\efa$. For a given switch epoch $m/H_*$, our procedure therefore requires no additional computation time relative to the usual switching procedure used by \textsc{AxionCAMB} which constructs an effective fluid from the instantaneous axion density at $m/H_*$. For a given accuracy, our procedure enables a massive improvement in computation time by enabling much earlier switches. 

We show in Fig.~\ref{fig:background_fluid_accuracy} the full end-to-end accuracy of our scheme at the background level by computing the axion density today for a fixed initial field displacement as a function of the switch epoch $m/H_*$. In the left panel, we compare our scheme (red) -- our specially constructed effective fluid and our appropriately chosen equation of state -- to the usual switching procedure and equation of state $w_{\CDM} = 0$ used by \textsc{AxionCAMB} (blue). The latter approach induces large oscillations in the axion density today  depending on the phase of the axion evolution picked out at $m/H_*$, leading to a large matching error which converges only as $(m/H_*)^{-1}$. On the other hand, our construction of the effective fluid eliminates the matching and evolution errors and allows even very early switch times $m/H_* = 10$ to reach better than $0.1\%$ accuracy. This $m/H_* = 10$ choice, marked with a star, is our reference solution scheme.

The right panel of Fig.~\ref{fig:background_fluid_accuracy} shows how our accuracy depends on the equation of state we choose, highlighting the importance of choosing the correct approximate equation of state to eliminate the evolution error. While our choice $w_\ax^\efa = (3/2) (m/H)^{-2}$ \eqref{eq:wax_next} performs so well that the dominant error of our scheme is the matching error from Fig.~\ref{fig:background_accuracy}, the choices $w_{\CDM}$ and $w_{\rm interp}$ yield much larger errors.

\begin{center}
  $\ast$~$\ast$~$\ast$
\end{center}

Our scheme for the background reaches accuracy better than $0.1\%$ in the final axion density with a switch to the EFA at $m/H_*=10$ --- before the axion field has even completed a full oscillation. We did this by constructing the effective fluid in such a way as to minimize the matching error between the exact effective fluid and the true axion at late times, and by evolving the effective fluid approximately using a better axion equation of the state than the ones commonly used in the literature. We now develop a similar scheme for the axion perturbations.

\subsection{Perturbations}
\label{subsec:perturbations}

The analysis of axion perturbations is more complicated than the background because they are continuously sourced by metric perturbations, and because the density perturbations have a Jeans scale below which axion density fluctuations oscillate rather than grow. These two effects are evident from the perturbed Klein-Gordon equation in synchronous gauge (e.g.\ Ref.~\cite{Hu:2004xd})
\begin{equation}
\label{eq:KG_pert}
\ddot{\delta\phi} + 2 \frac{\dot a}{a} \dot{\delta\phi} + (k^2+a^2 m^2) \delta\phi = - \frac{\dot h_L}{2} \dot\phi\,,
\end{equation}
which looks like the background equation \eqref{eq:bgKG} but for the new $k^2$ term and the sourcing by the time derivative of $h_L$, the trace of the spatial metric perturbation. 

We first examine the unsourced homogeneous left-hand side of Eq.~(\ref{eq:KG_pert}) in radiation domination. The frequency $k^2 + a^2 m^2 - \ddot{a}/a$ now implies oscillations for sufficiently large $k$ even  after averaging over the mass induced oscillations. These correspond to acoustic or Jeans oscillations in the effective fluid.

We can better understand the two oscillation scales by formally extracting the $m t=\tau$ term from the argument of the cosine when we write the WKB solution, obtaining the leading order solution in the form 
\begin{equation}
\label{eq:WKBpert_split}
\delta \phi \propto \cos\left[\tau +  k \int c_{s\phi} d\eta + \alpha\right]\,,
\end{equation}
where $\alpha$ here is a phase and the sound speed for field fluctuations is
\begin{equation}
\label{eq:cs2_field}
c_{s\phi} = \left(\frac{k}{a m}\right)^{-1} \left( \sqrt{1+\left(\frac{k}{a m} \right)^2} -1 \right) \,.
\end{equation}

Now if we expand out the cosine into $\cos(\tau)$ and $\sin(\tau)$ we can see that even after averaging over the $\tau$ scale there remain field oscillations which are then imprinted into the axion density perturbations.

The sound speed for field fluctuations approaches $c_{s\phi} = 1$ at early times $k / a m \gg 1$ and $c_{s\phi} = k/2 a m$ at late times $k/ a m \ll 1$. Since $\eta\propto a$ during radiation domination and $\propto a^{1/2}$ during matter domination, we can see that for $m\gg \Heq$ the net effect of the sound speed at late times is captured by the contributions around matter radiation equality $\aeq$,
\begin{equation}
k \int_0^{\eta_{0}} c_{s\phi} d\eta \propto \frac{k^2}{\aeq^2 m \Heq}\,. 
\end{equation}
It is therefore useful to define the maximal Jeans scale as 
\begin{equation}
k_J\equiv 3^{1/4} \aeq \sqrt{ m \Heq}\,,
\end{equation}
where the numerical factor corresponds to the choice in the literature and is based on its impact on the effective fluid \cite{Hu:2000ke}. 

This impact can be clearly seen by evaluating the density, pressure and divergence of momentum density perturbations
\begin{align*}
\label{eq:sources}
\delta \rho_\ax &= a^{-2} \dot\phi \dot{\delta\phi} + m^2 \phi \delta\phi\,,\\
\delta P_\ax & = a^{-2} \dot\phi \dot{\delta\phi} - m^2 \phi \delta\phi\,, \\
(\rho_\ax +P_\ax)\theta_\ax &= a^{-2} k^2 \dot\phi \delta\phi\,, \numberthis
\end{align*}
using the WKB solution \eqref{eq:WKBpert_split} and then averaging over the mass oscillations to find
\begin{equation}
\label{eq:cs2deltaPdeltarho}
 \frac{\langle \delta P_\ax \rangle}{\langle \delta \rho_\ax \rangle} \simeq c_{s\phi}^2\,,
\end{equation}
where the field sound speed plays the role of a sound speed for the fluid and the approximation is leading order in $(m/H)^{-1}$. Under the Jeans scale, pressure fluctuations therefore support the axion density perturbations against gravitational collapse. Given that this suppression of axion density perturbation growth relative to CDM is the 
main effect of the axion mass, we seek to quantify the suppression out to $k\sim k_J$ using efficient but accurate techniques similar to those we introduced for the background. 

We follow the same procedure as for the background. We start at some $m/H_*$ by splitting the axion perturbation $\delta \phi$ into two pieces $\delta\varphi_{c,s}$ as 
\begin{align*}
\label{eq:pert_ansatz}
\delta\phi(\tau) &= \delta\varphi_c(\tau) \cos\left[\tau - \tau_*  \right] + \delta\varphi_s(\tau) \sin\left[\tau - \tau_* \right] \numberthis\,.
\end{align*}

The equations of motion for $\delta \varphi_{c,s}$ follow from plugging the background ansatz \eqref{eq:ansatz} and the perturbation ansatz \eqref{eq:pert_ansatz} into the perturbed Klein-Gordon equation \eqref{eq:KG_pert} and again imposing that the cosine and sine pieces are solved separately, yielding respectively
\begin{align*}
\label{eq:pertKGvarc}
\delta \varphi_c''  +  2 \delta \varphi_s' &+ 3 \frac{H}{m} \left( \delta \varphi_c' + \delta \varphi_s \right) \numberthis \\ &+ \frac{k^2}{a^2 m^2} \delta \varphi_c = -\frac{h_L'}{2} \left(\varphi_c'+ \varphi_s \right)\,,
\end{align*}
and
\begin{align*}
\label{eq:pertKGvars}
\delta \varphi_s''-2 \delta \varphi_c'&+ 3 \frac{H}{m} \left(- \delta \varphi_c + \delta \varphi_s'\right) \numberthis \\&+ \frac{k^2}{a^2 m^2} \delta \varphi_s = -\frac{h_L'}{2} \left(\varphi_s'- \varphi_c \right)\,. 
\end{align*}

The first line of each of these equations looks like the background $\varphi_{c,s}$ equations \eqref{eq:bgKGvar}. The second line contains the new $k/a m$ dependent term and the metric source. We see that the metric source in the equation of motion is multiplied by coefficients $\left(\varphi_c'+ \varphi_s \right)$ and $\left(\varphi_s'- \varphi_c \right)$. While these background quantities evolve as $a^{-3/2}$ in the absence of background matching error, they also transfer their fast oscillatory matching error \eqref{eq:bgsol} to the perturbations.

We then construct from $\delta \varphi_{c, s}$ the effective fluid versions of the quantities in Eq.~(\ref{eq:sources})
\begin{align*}
\label{eq:fluidperts}
\delta \rho_\ax^\ef   ={}&\ \frac{1}{2} m^2 \left[\varphi_s \delta\varphi_c' - \varphi_c \delta\varphi_s' + \delta\varphi_c' \varphi_c'+ \delta\varphi_s' \varphi_s' \right.\\&+ \left.  \delta\varphi_s (2 \varphi_s + \varphi_c') + \delta\varphi_c  (2 \varphi_c - \varphi_s') \right]\,,\\
\delta P_\ax^\ef  ={}&\ \delta\rho_\ax^\ef - m^2 \left[\delta \varphi_s \varphi_s + \delta \varphi_c \varphi_c \right]\,,\numberthis\\
(\rho_\ax^\ef  + P_\ax^\ef)\theta_\ax^\ef ={}&\ \frac{k^2 m}{2 a} \left[\delta \varphi_c \left(\varphi_s+\varphi_c'\right)+ \delta \varphi_s\left(-\varphi_c+\varphi_s'\right)\right]\,,
\end{align*}
in such a way that they satisfy the conservation equations for the effective fluid in synchronous gauge \cite{Hu:2004xd}
\begin{align*}
\frac{d \delta \rho_\ax^\ef}{d \eta} + 3 {\frac{\dot{a}}{a}} \left(\delta \rho_\ax^\ef + \delta P_\ax^\ef\right) &= -(\rho_\ax^\ef + P_\ax^\ef) (\theta_\ax^\ef +\frac{1}{2} \dot{h}_L)\,, \\
\left[\frac{d}{d\eta} + 4 {\frac{\dot{a}}{a}}\right] \left(\rho_\ax^\ef+P_\ax^\ef \right) \frac{\theta_\ax^\ef}{k^2} &= \delta P_\ax^\ef\,. \numberthis
\label{eq:perteom}
\end{align*}

Just like the method for the background, the effective fluid simply recasts the axion perturbations in different variables with no approximations. We again call a matching error the difference between $\delta\rho_\ax$ and $\delta\rho_\ax^\ef$ at late times when $m/H \gg 1$.   Likewise since Eq.~\eqref{eq:perteom} are the equations of motion for an effective fluid but require a closure condition to define $\delta P_\ax^\ef$, we seek to approximate it by calibrating the equation of state for the perturbations, i.e.\ the sound speed, so that we no longer need to solve for the auxiliary variables after the switch.  Specifically, the conservation equations are equivalent in the generalized dark matter language to effective fluid equations of motion \cite{Hu:1998kj}
{\begin{align*}
\label{eq:ef_pert}
\dot{\delta}_\ax^\efa ={}& -\left(1 + w_\ax^\efa \right) \left(\theta_\ax^\efa + \frac{\dot{h}_L}{2}\right)  - 3 \left(c_s^2 -w_\ax^\efa \right) {\frac{\dot{a}}{a}} \delta_\ax^\efa \\&- 9\left(1+ w_\ax^\efa \right) \left(c_s^2 - c_a^2\right) \left(\frac{\dot{a}}{a}\right)^2 \frac{\theta_\ax^\efa}{k^2}\,, \\
\dot{\theta}_\ax^\efa ={}& -(1- 3 c_s^2) {\frac{\dot{a}}{a}} \theta_\ax^\efa + \frac{c_s^2 k^2}{1 + w_\ax} \delta_\ax^\efa\,, \numberthis
\end{align*}
where $\delta_\ax^\efa \equiv \delta \rho_\ax^\efa/\rho_\ax^\efa$, the adiabatic sound speed $c_a$ is determined by $w_\ax^\efa = P_\ax^\efa /\rho_\ax^\efa$ as
\begin{equation}
\label{eq:ca2}
c_a^2 \equiv \frac{\dot{P}_\ax^\efa}{\dot{\rho}_\ax^\efa} = w_\ax^\efa - \frac{\dot w_\ax^\efa}{3 ( 1 + w_\ax^\efa) {\frac{\dot{a}}{a}}}\,,
\end{equation}
and $c_s^2$ is the sound speed of the effective fluid in its rest frame,
\begin{equation}
\label{eq:cs2_fluid}
c_s^2 \equiv \left.\frac{\delta P^\efa_\ax}{\delta \rho^\efa_\ax}\right\vert_{\rm rest}\,,
\end{equation}
where the rest frame can be accessed through the gauge transformation
\begin{equation}
\eta \rightarrow \eta + \theta_\ax^\efa/k^2\,,
\end{equation}
yielding
\begin{align*}
\label{eq:gauge_transformations}
\left.\delta \rho_\ax^\efa\right\vert_{\rm rest} &= \left.\delta \rho_\ax^\efa + 3 {\frac{\dot{a}}{a}} (\rho_\ax^\efa + P_\ax^\efa) \frac{\theta_\ax^\efa}{k^2} \right\vert_{\rm sync}\,,\\
\left. \delta P_\ax^\efa \right\vert_{\rm rest} &= \left. \delta P_\ax^\efa +   3 {\frac{\dot{a}}{a}}  c_a^2 (\rho_\ax^\efa + P_\ax^\efa) \frac{\theta_\ax^\efa}{k^2} \right\vert_{\rm sync}\,. \numberthis
\end{align*}
For notational simplicity, we have dropped the ``efa" marker on $c_s^2$ here and restore it below where confusion might arise. Just as in the background case, we call the error induced by employing the effective fluid approximation to close the system with $w_\ax^\efa$ and $c_s^2$ an evolution error. Again we analyze the matching and evolution errors in turn. 

\subsubsection{error}
\label{subsubsec:matching_pert}

In the absence of the Jeans oscillations and the metric sourcing described above, the perturbation equations take the same form as the background equations. We therefore choose the same matching conditions (\ref{eq:constraints}) as the background
\begin{align}
\label{eq:pert_constraints}
\left.\frac{\delta \varphi_{c,s}''}{\delta \varphi_{c,s}'}\right\vert_{*} = \left.-\frac{1}{2} \frac{\langle H \rangle}{m} \left(3 - \frac{m}{\langle H\rangle ^3}\frac{d\langle H\rangle^2}{d \tau}\right)\right\vert_{*}\,,
\end{align}
and quantify the additional error induced by the new effects.

\begin{figure}[t]
	\includegraphics{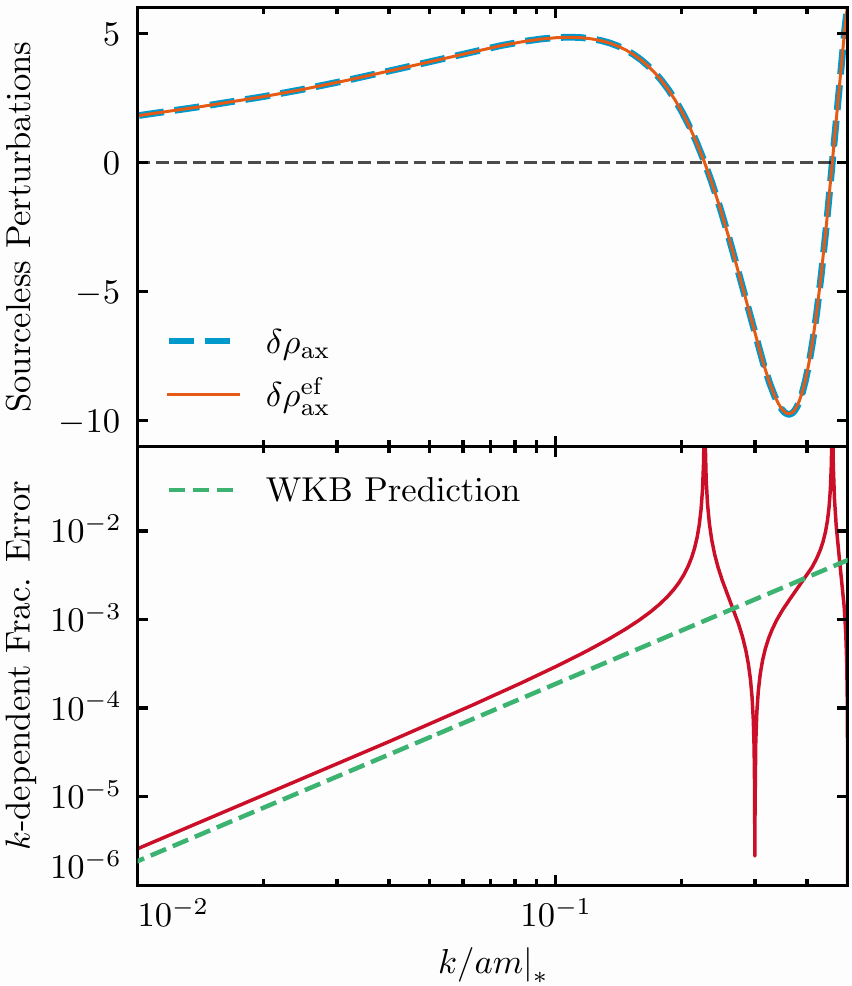}
	\caption{Without metric sourcing, the effective fluid density perturbation $\delta \rho_\ax^\ef$ (top panel, orange) makes a small $k/a m\vert_*$-dependent matching error (bottom panel, red) relative to the true axion density $\delta \rho_\ax$ (top panel, blue dashed) at late times. We fix here the switch epoch $m/H_* = 10$ so that the $k$-independent $m/H_*$ contribution to the error can be subtracted out. Each point thus represents the error for a different $k$-mode. Using the WKB approximation we find in Eq.~\eqref{eq:kam_matching_error_WKB} that this error goes as $3/16 \times (m /H)^{-1} \times (k/(a m))^2 \vert_*$ times a phase factor (bottom panel, green dashed), which is confirmed by the numeric solution (neglecting metric sourcing).  Poles in the error are due simply to zero-crossings of the density perturbation as seen in the top panel. Metric sourcing further reduces this error in practice. For further discussion see \S\ref{subsubsec:matching_pert}.}
	\label{fig:kam_matching_error_sourceless}
\end{figure}

Let us first understand the additional effect of Jeans oscillations on the matching error in the absence of metric sourcing. In this case the WKB solution (\ref{eq:WKBpert_split}) holds and we can take a background solution with no matching error $A=B=0$ to assess the additional matching error produced by Eq.~(\ref{eq:pert_constraints}). As with the background these take a form dictated by the trigonometric identities (\ref{eq:trig}) and we can solve for their amplitudes $\delta A$, $\delta B$ given the matching conditions \eqref{eq:pert_constraints}. From these we infer an additional matching error from the Jeans oscillations that does not diminish with time and scales with the matching epoch as
\begin{align}
\label{eq:kam_matching_error_WKB}
\frac{(\delta \rho_\ax^\ef - \delta\rho_\ax)}{\delta\rho_\ax} & \supset \left.\frac{3}{16} {\left(\frac{m}{H}\right)^{-1}} \left(\frac{k}{a m}\right)^2 \right\vert_* \times \textrm{phase} \,,
\end{align}
for $k/a m\vert_* \lesssim 1$. Here ``phase" designates an ${\cal O}(1)$ coefficient that depends on the phase of the oscillations in the background and perturbations at $m/H_*$.  

We confirm this analytic result for the Jeans-induced additional matching error in Fig.~\ref{fig:kam_matching_error_sourceless} by solving the system numerically with metric sourcing turned off by hand. We quantify the matching error in the same way we did for the background. Since it must vanish for a very late switch $m/H_* \gg 1$, we can compute it by comparing $\delta \rho_\ax^\ef$ from an early switch, here $m/H_* = 10$, to a much later switch. We change $(k/a m)\vert_*$ by changing $k$. By keeping $m/H_*$ fixed, we can subtract off the non-$k$ dependent piece of the matching error and we find that the $k$-dependent error agrees well with the analytic result \eqref{eq:kam_matching_error_WKB}.

Notice that for $m/H_*=10$ this error remains ${\cal O}(10^{-3})$ even for scales that have already undergone a couple of Jeans oscillations and hence are at most comparable to the ${\cal O}(m/H_*)^3$ errors we induce in their absence. Poles in the error are due only to zero crossings of the density perturbations (top panel) and do not reflect an increase in the absolute error of our scheme.  We shall see that on these scales the final transfer function has already been so suppressed relative to CDM that higher accuracy is not required in practice.

Next let us consider the effect of metric sourcing on the perturbations assuming that the Jeans corrections are small $k/a m \ll 1$. The metric perturbations satisfy the Einstein momentum and Poisson equations in synchronous gauge
\begin{align*}
\dot{\eta}_T &= \frac{1}{2} a^2 \left( \rho + P \right) \frac{v}{k}\,, \\
-k^2 \eta_T + \frac{1}{2} \frac{\dot{a}}{a} \dot{h}_L &= \frac{1}{2} a^2 \delta \rho\,, \numberthis
\end{align*}
where $\eta_T$ is the curvature perturbation. $h_L$ and $\eta_T$ are initialized in the superhorizon radiation-dominated regime and are determined by radiation perturbations which take their usual form
(see, e.g., Refs.~\cite{Ma:1995ey,Lin:2018nxe}). In this work, we include massless neutrinos in the radiation component as a perfect fluid, neglecting their anisotropic stress and so during radiation domination
\begin{equation}
\dot h_L \propto
\begin{cases}
a \quad &\textrm{(superhorizon)}\,,  \\
a^{-1} \quad &\textrm{(subhorizon)}\,.
\end{cases}
\end{equation}
When combined with the scaling of field fluctuations in the background $\phi  \propto a^{-3/2}$,
these sources then continuously generate  field perturbations through Eq. (\ref{eq:KG_pert}) that scale as
\begin{equation}
\label{eq:field_sourcing}
\delta \phi  \propto
\begin{cases}
{a^{1/2}} \quad &\textrm{(superhorizon)}\,,  \\
a^{-3/2} \ln(\tau)  \quad &\textrm{(subhorizon)}\,.
\end{cases}
\end{equation}
where the $\ln(\tau)$ term corresponds to the familiar logarithmic growth of matter density fluctuations during radiation domination.  

\begin{figure}[t]
	\includegraphics{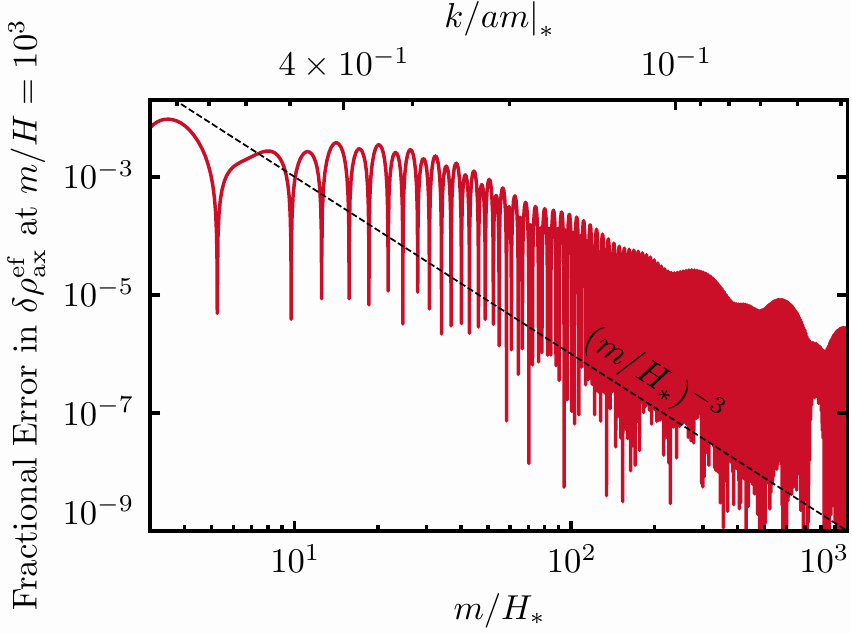}
	\caption{With metric sourcing, $\delta \rho_\ax^\ef$ for $k=k_J$ evaluated at a fixed time $m/H = 10^3$ is already $\sim0.1\%$ converged as a function of the switch epoch even for early switch epochs $m/H_* \sim 10$. The convergence rate is slower than the simple $\mathcal{O}(m/H_*)^{-3}$ estimate we derived analytically in the $k\rightarrow0$, $H \ll H_*$, $\rho_\ax \rightarrow 0$ limit. For further discussion see \S\ref{subsubsec:matching_pert}.}
	\label{fig:perturbation_accuracy}
\end{figure}

This metric sourcing has two types of effects on the matching error.   
First it alters the optimal matching coefficient so that Eq.~\eqref{eq:pert_constraints} no longer produces the full
$(m/H_*)^{-3}$ suppression of errors as it does in the background. However, the metric sourcing continues after the matching and so to the extent that it dominates the final field perturbation, the perturbation matching error itself goes away. In the superhorizon limit for matching, the strong growth of the field fluctuation implies a substantial mitigation of perturbation matching errors whereas in the subhorizon limit this mitigation is only logarithmic. This mitigation also applies to the matching error from the Jeans term, Eq.~\eqref{eq:kam_matching_error_WKB}. In radiation domination we have
\begin{align*}
\left. \frac{k}{a H}\right\vert_{*} &\sim \frac{k}{k_{J}} \left(\frac{m}{H_*}\right)^{1/2}\,, \numberthis
\end{align*}
such that for the highest $k \sim k_J$ modes that we are interested in and for $m/H_* \sim 10$, the switch occurs just under the horizon and the mitigation for these modes is between power law and logarithmic. For much smaller $k \ll k_J$ modes the power law mitigation makes the perturbation matching error entirely irrelevant.

Second, even if these perturbation matching errors from  Eq.~(\ref{eq:pert_constraints}) go away due to sourcing, any background matching error will regenerate an error through the $h_L' \varphi_{s,c}$ terms in the perturbation equations of motion Eqs.~(\ref{eq:pertKGvarc}) and (\ref{eq:pertKGvars}), leaving an unavoidable ${\cal O}(m/H_*)^{-3}$ fractional error in $\rho_\ax^\ef$ at the end.

In Fig.~\ref{fig:perturbation_accuracy}, we estimate the full matching error for the largest wavenumber of interest $k = k_J$, and hence the largest matching error, numerically. This mode has $k/a m \sim 0.5$ at $m/H_* = 10$. We estimate the matching error using the same procedure we used for the background in Fig.~\ref{fig:background_accuracy}: we track convergence of the effective fluid $\delta \rho^\ef$ evaluated at a fixed time $m/H = 10^3$ as the switch epoch $m/H_*$ is taken later and later. 

\begin{figure*}
	\includegraphics{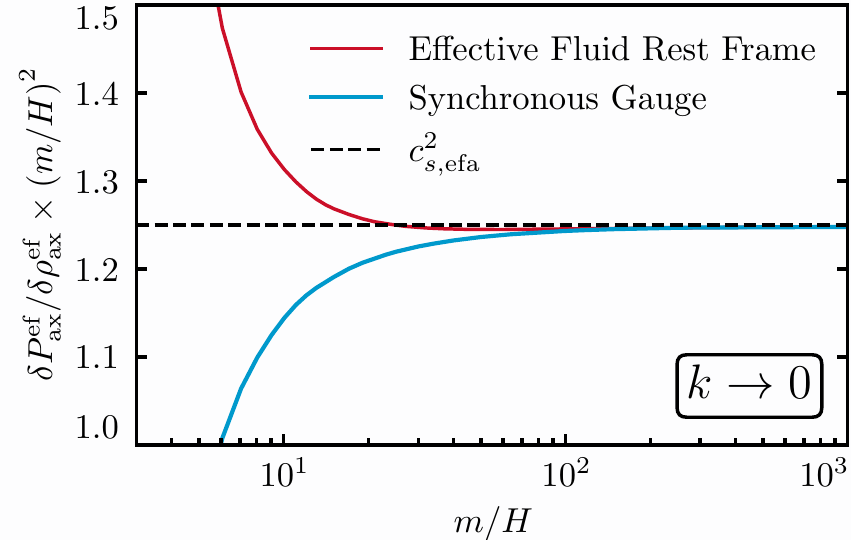}
	\includegraphics{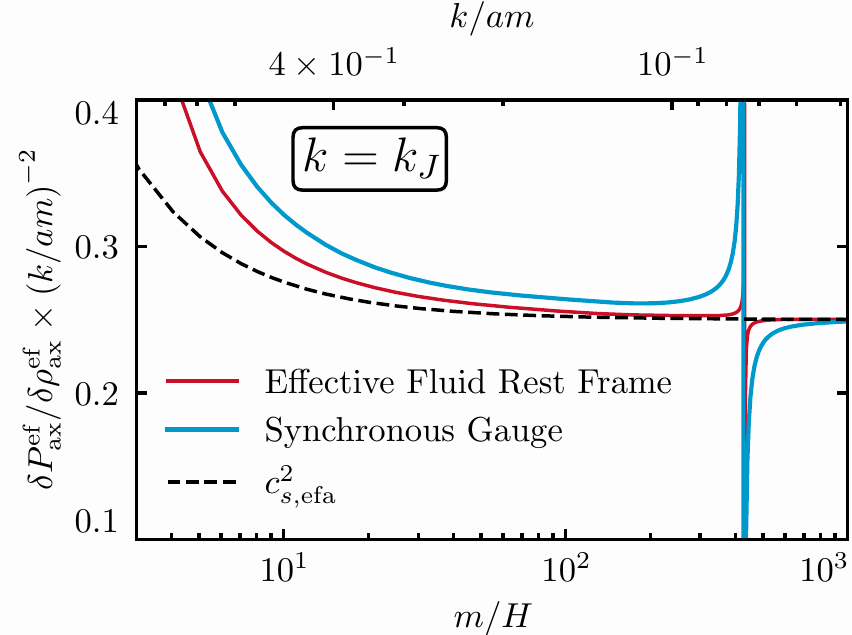}
	\caption{Left: The effective fluid sound speed $\delta P_\ax^\ef/\delta \rho_\ax^\ef$ in the $k\rightarrow0$ limit, multiplied by $(m/H)^2$. The result in the rest frame (red) extracts the amplitude of the leading order $(m/H)^2$ term in the fluid sound speed, which we use to improve the field-derived $c_{s\phi}$ when we construct the effective fluid approximation $c_{s, \efa}^2$. Right: The effective fluid sound speed $\delta P_\ax^\ef/\delta \rho_\ax^\ef$ for  $k=k_J$, multiplied by the leading order scaling $(k/a m)^{-2}$. Our effective fluid approximation $c_{s, \efa}^2$ is a good approximation for $\delta P_\ax^\ef/\delta \rho_\ax^\ef$ in the rest frame. In both figures, the effective fluid is constructed at the time $m/H_* = m/H$ for each point on the solid curves.  For further discussion see \S\ref{subsubsec:evolution_pert}.
	}
	\label{fig:cs2}
\end{figure*}

The perturbation matching error for $m/H_* = 10$ is slightly larger than $\sim0.1\%$, which is marginally larger than the background error. Our error improves as $m/H_*$ and $k/a m\vert_*$ become insignificant, though significantly more slowly than $(m/H_*)^{-3}$ mainly due to the Jeans scale  $\mathcal{O}(k/a m\vert_*)$ effects described above.

The overdensity $\delta_\ax^\ef $ and the velocity perturbation $\theta_\ax^\ef $ involve dividing these perturbed effective fluid quantities by the background effective fluid quantities $\rho_\ax^\ef$ and $\rho_\ax^\ef + P_\ax^\ef$. Since our scheme is more accurate for the background than for the perturbations, the matching errors we have studied here are the dominant matching errors in $\delta_\ax^\ef$ and $\theta_\ax^\ef$. However, in the $k\rightarrow0$ limit the matching errors for the perturbations and the background become identical and the errors are in phase, such that the errors cancel in $\delta_\ax^\ef$ and $\theta_\ax^\ef$ which become more accurate than their separate components.

\subsubsection{Evolution error}
\label{subsubsec:evolution_pert}

Just as in the case of the background, we want to approximate the effective fluid conservation law \eqref{eq:ef_pert} by replacing $\delta P_\ax^\ef/\delta \rho_\ax^\ef\vert_{\rm rest}$ \eqref{eq:cs2_fluid} with an approximate equation of state $c_{s, \efa}^2$.

At late times the sound speed goes to zero, but it should have corrections for finite $k/am$ and $(m/H)^{-1}$. The leading order $k/am$ type corrections are encapsulated by the field sound speed $c_{s\phi}^2$ \eqref{eq:cs2_field}. In the absence of metric sourcing and the $k/am$ type oscillations, the leading order $(m/H)^{-1}$ type corrections would be the same as the background $w_\ax \sim 3/2 (m/H)^{-2}$. However we find a deviation from this behavior.

In the left panel of Fig.~\ref{fig:cs2}, we plot the exact $\delta P_\ax^\ef/\delta \rho_\ax^\ef$ for a very large-scale mode $k\ll k_J$. For such a mode, $k/a m$ type effects are negligible at late times $m/H \gtrsim 1$ and so the field sound speed $c_{s \phi}$ goes to zero, but $(m/H)^{-1}$ effects can still be significant. We can solve our auxiliary Klein-Gordon equations  \eqref{eq:pertKGvarc} and \eqref{eq:pertKGvars} for our $\delta \varphi_{c,s}$ auxiliary variables, compute $\delta \rho^\ef$ and $\delta P^\ef$ in synchronous gauge, and then use the gauge transformations Eq.~\eqref{eq:gauge_transformations} to access their values in the effective fluid rest frame, from which we can compute the sound speed.  Just as in the case of the background we do not attempt to fit the $m/H_*$ dependent piece of the sound speed and therefore we minimize it by setting $m/H_* = m/H$ for each time in the figure.

By doing so we can see directly that the effective sound speed is not zero as $k/a m \rightarrow 0$ but instead exhibits a $\mathcal{O}(m/H)^{-2}$ type correction as we expected. However the coefficient of this $(m/H)^{-2}$ term is $\sim5/4$ rather than $\sim3/2$ as might have been naively guessed from the study of the background. This shows a key benefit of our effective fluid approach -- it enables us to self-calibrate the effective fluid approximation more effectively than we might have been able to with analytics alone.

We therefore choose the EFA sound speed
\begin{equation}
\label{eq:cs2_next}
c_{s,\efa}^2 = c_{s\phi}^2 + \frac{5}{4} \frac{H^2}{m^2}\,,
\end{equation}
which encompasses the leading order $k/a m$ and $(m/H)^{-1}$ corrections to the asymptotic limit $c_s^2 \rightarrow 0$.

In the right panel of Fig.~\ref{fig:cs2}, we show that our EFA sound speed is a good approximation for the sound speed of the effective fluid for a large $k$-mode $k = k_J$.  While the asymptotic behavior is set by the $(k/am)^2$ behavior of the field sound speed $c_{s\phi}^2$, at $m/H=10$ both pieces of our EFA sound speed are important to successfully approximate $c_{s}$. While there is a small difference between the effective fluid sound speed and our approximation $c_{s, \efa}^2$ at $m/H = 10$, the sound speed is relatively small at this stage ($\sim 0.05$) which suppresses the effect of this small error on the density perturbations. 

The pole in the right-hand panel of Fig.~\ref{fig:cs2} corresponds to a zero crossing in the effective fluid density perturbation $\delta \rho_\ax^\ef$ in synchronous gauge (see the yellow line in Fig.~\ref{fig:graphic}). The gauge transformation to reach the effective fluid rest frame is highly oscillatory and therefore the density in that gauge $\delta \rho_\ax^\ef\vert_{\rm rest}$ \eqref{eq:gauge_transformations} also has a nearby zero. The synchronous gauge pressure perturbation $\delta P_\ax^\ef$ oscillates and therefore the sound speed in synchronous gauge $\delta P_\ax^\ef / \delta \rho_\ax^\ef$ has a pole. In the rest frame, however, $\delta P_\ax^\ef\vert_{\rm rest}$ does not oscillate and instead has a single zero crossing at nearly, but not exactly, the same time as $\delta \rho_\ax^\ef\vert_{\rm rest}$. This indicates that $\delta P_\ax^\ef\vert_{\rm rest}$ has a very small component which is not proportional to $\delta \rho_\ax^\ef\vert_{\rm rest}$ and therefore cannot be exactly modeled by a sound speed (see Eq.~\eqref{eq:cs2_fluid}). This transient component scales as $(H/m) (k/am)^2$ and is small enough that it does not significantly impact the evolution of the perturbations of interest. 

\begin{figure}[t]
\includegraphics[]{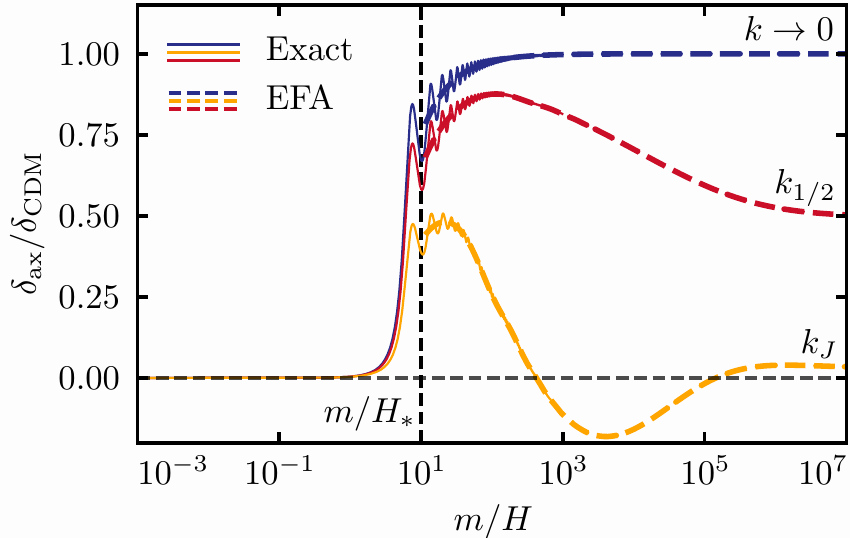}
\caption{Our reference solution scheme for the axion perturbations. At early times $m/H < m/H_* = 10$, we solve the exact (solid lines) Klein-Gordon equations of motion for the axion field. We then switch to the EFA (dashed lines), with our auxiliary variables providing matching conditions at the switch. These matching conditions allow the EFA to effectively act as a cycle-averaged axion and thus match the true axion at late times. Evolution in the EFA is performed with our optimized equation of state and sound speed. At late times, axion perturbations are unsuppressed relative to CDM on large scales (purple) but suppressed when $k\sim k_J$ (yellow). The scale $k_{1/2}$ (red) is defined by $\delta_\ax/\delta_\CDM (k_{1/2}) \equiv 1/2$ at late times. Complete axion transfer functions are shown in Fig.~\ref{fig:Tk}, and for further discussion see \S\ref{subsubsec:evolution_pert}.}
\label{fig:graphic}
\end{figure}

With the EFA sound speed now appropriately defined, we finally have a complete effective fluid approximation for the axion. We summarize it in Fig.~\ref{fig:graphic} for a range of relevant $k$. This figure is the perturbation parallel to the bottom panel of the background Fig.~\ref{fig:background}. After solving for the usual Klein-Gordon equation for the axion field perturbation in the Hubble drag $m \lesssim H$ regime, we switch to the EFA at $m/H_* = 10$.

Before the switch, we compute and show the true axion density perturbation $\delta_\ax$. After the switch, we compute and show the effective fluid approximation for the density perturbation $\delta_\ax^\efa$. These are discontinuous at the switch time because the effective fluid approximation is constructed to match the true axion at late times, rather than at the matching point.

After evolving to late times in the EFA, the large-scale mode $k\rightarrow0$ shows no suppression of the axion perturbations $\delta_\ax$ relative to the CDM overdensity $\delta_{\CDM}$. For the Jeans scale $k= k_J$, on the other hand, the suppression is significant.

We define the mode $k_{1/2}$ where the axion density perturbation today relative to CDM reaches one half, since this mode represents a point where the suppression is substantial, but the linear theory power remains appreciable and therefore represents a convenient location to benchmark accuracy. This is a slightly larger scale than the Jeans scale, with $k_{1/2} \simeq 0.54 \ k_J$ for a $10^{-22}$ eV axion. 

\begin{figure}[t]
\includegraphics[]{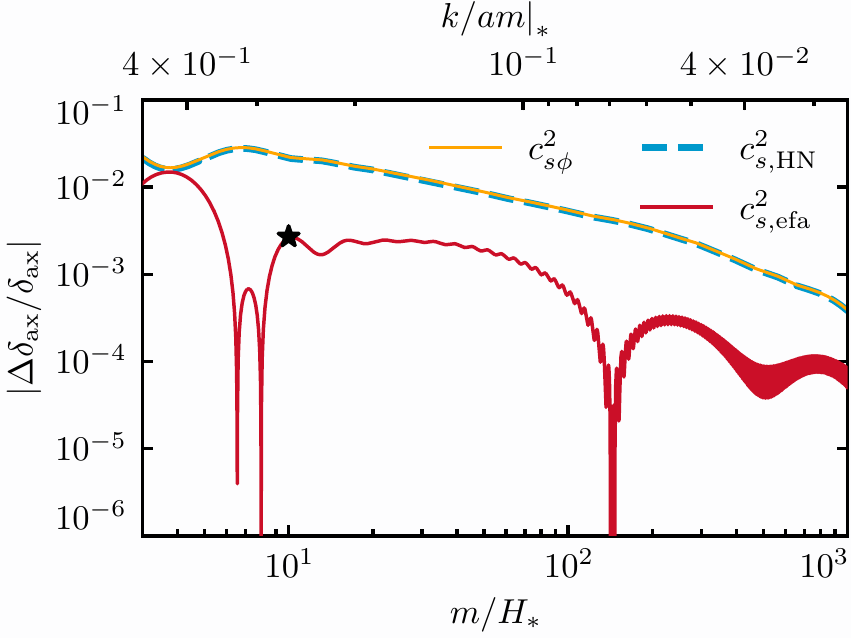}
\caption{For a mode $k_{1/2}$ where the matter power spectrum yields half its CDM value, we show our technique's total error as a function of our switch time parameter $m/H_*$ for various choices of the EFA equation of state $c_s^2$, computed by comparing to a late switch time $m/H_* = 2 \times 10^3$. Using the field sound speed $c_{s\phi}$ \eqref{eq:cs2_field} or the sound speed $c_{s, \textrm{HN}}$ \eqref{eq:cs2_HN} of Hwang \& Noh (2009) \cite{Hwang:2009js}, our approach makes a few percent error for a switch time $m/H_*=10$. Our EFA sound speed \eqref{eq:cs2_next} resolves the bulk of this error and enables us to achieve a subpercent error with a switch time of $m/H_*=10$, marked by a star. For further discussion see \S\ref{subsubsec:evolution_pert}.}
\label{fig:pert_fluid_accuracy}
\end{figure}

We test the full accuracy of our scheme at $k_{1/2}$ in Fig.~\ref{fig:pert_fluid_accuracy}, including the matching errors from \S\ref{subsubsec:matching_pert} and the evolution errors induced by replacing $c_s^2$ with $c_{s, \efa}^2$.  We evaluate $\delta_\ax^{\efa}$ at late times ($k/a m = 10^{-4}$) and check its dependence on the switch time $m/H_*$. We see that with our choice of EFA sound speed we already reach a subpercent accuracy at $m/H_* = 10$ for $k_{1/2}$. 

If we had used only the field sound speed $c_{s \phi}$ \eqref{eq:cs2_field} to approximate the effective fluid sound speed, we would have made a much larger evolution error of several percent.
An alternative sound speed often used in the literature is derived in Hwang \& Noh (2009) \cite{Hwang:2009js},
\begin{equation}
\label{eq:cs2_HN}
c_{s, \textrm{HN}}^2
\simeq \frac{k^2}{4 a^2 m^2 + k^2}\,,
\end{equation}
which though it has the same limits as our field sound speed $c_{s \phi}$ differs at order $k/a m$.  It also does not include the $\mathcal{O}(m/H)^{-2}$ correction of our $c_{s, \rm efa}^2$, and therefore as shown in Fig.~\ref{fig:pert_fluid_accuracy} its error properties are similar to those of $c_{s \phi}$, leading to a much larger evolution error than our $c_{s, \efa}^2$.

We show in Fig.~\ref{fig:k_accuracy} that the choice of switch epoch $m/H_* = 10$ yields sufficient accuracy throughout the range of scales $k$ and axion masses $m$ in which we are interested by comparing the axion density perturbation $\delta_\ax$ computed with a late switch $m/H_*=1000$ to our reference $m/H_* = 10$ scheme. To show different mass axions with on the same axes, we scale the horizontal axis by the mass-dependent $k_{1/2}$. 

The accuracy of our choice $m/H_* = 10$, shown in the bottom panel, is well behaved as a function of $k$ at a fixed mass. The pole here corresponds to the node in the density perturbation, as shown in the top panel -- the absolute error remains small throughout. As a function of mass at fixed $k/k_{1/2}$, we see that our accuracy increases for heavier axions and decreases for lighter axions. This reflects that all our scalings are tuned to work best when the switches occur deep in radiation domination.  Nonetheless for the half-amplitude mode $k_{1/2}$ our parameter choice $m/H_*=10$ yields subpercent accuracy for all masses shown.

\begin{figure}[t]
\includegraphics[]{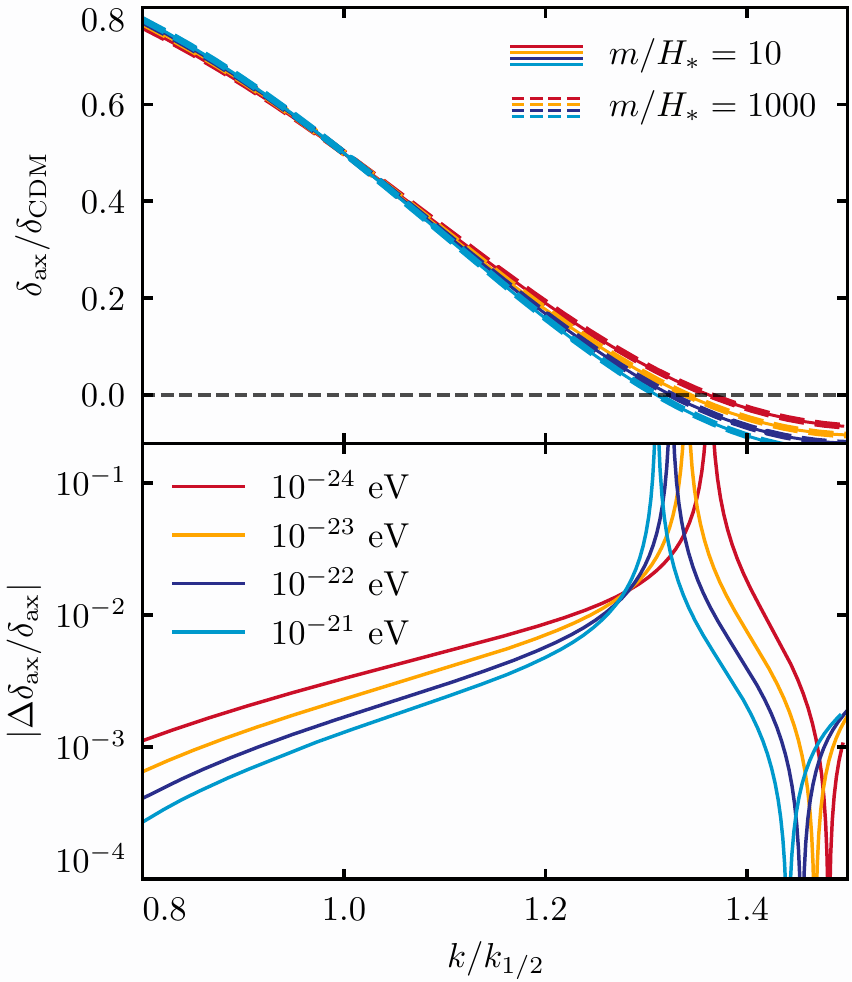}
\caption{Our effective fluid procedure accurately tracks the decline of the axion power spectrum relative to CDM as a function of scale $k$ for all relevant masses (top panel), with our reference $m/H_*=10$ scheme (solid lines) visually indistinguishable from a more accurate result which uses a very late switch time $m/H_*=1000$ (dashed lines). The fractional difference in the bottom panel represents the accuracy of our reference scheme, with the pole here simply due to the node in the transfer function. The horizontal axis is scaled by $k_{1/2}$, the mass-dependent scale defined by $\delta_\ax/\delta_\CDM (k_{1/2}) \equiv 1/2$. For further discussion see \S\ref{subsubsec:evolution_pert}.}  
\label{fig:k_accuracy}
\end{figure}

\section{Results}
\label{sec:results}

\begin{figure*}[t]
\includegraphics[]{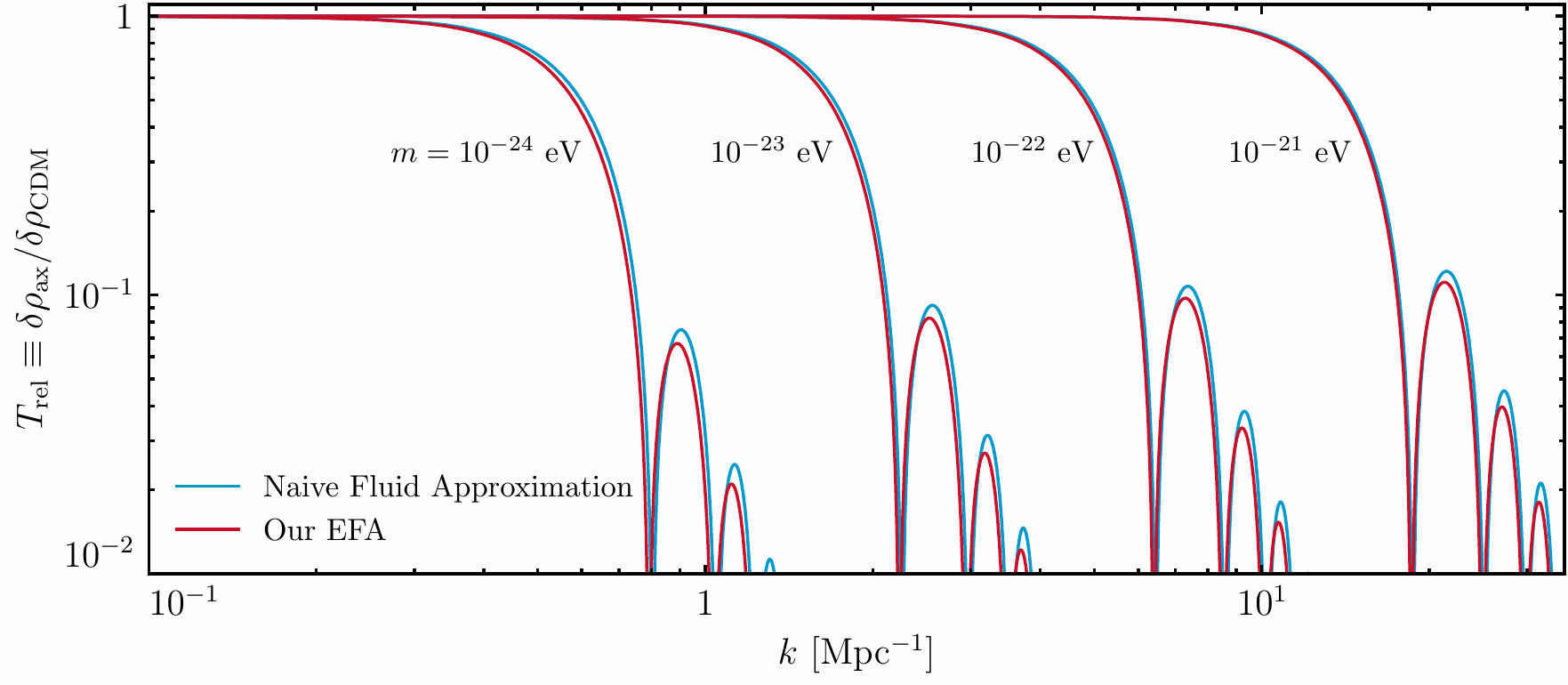}
\caption{The axion transfer function relative to CDM, Eq.~\eqref{eq:Trel}. The method developed in this work (red) with switch parameter $m/H_* = 10$ yields subpercent accuracy for the half-amplitude mode $k_{1/2}$ where $T_{\rm rel} (k_{1/2}) = 1/2$ (see \S\ref{sec:method} for a detailed error characterization). The naive fluid approach (blue) has no significant computational advantage over our approach but underestimates the power spectrum cutoff induced by the axion by as much as several percent, an error which we quantify in Fig.~\ref{fig:fluid_acc} and Fig.~\ref{fig:mass_shift}. For further discussion see \S\ref{sec:results}.} 
\label{fig:Tk}
\end{figure*}

We now use our solution scheme to compute the axion transfer function relative to CDM of the same density today,
\begin{equation}
\label{eq:Trel}
T_{\rm rel}(k) \equiv  \frac{\delta \rho_\ax (k,{a=1})}{\delta \rho_{\CDM} (k,{a=1})}\,,
\end{equation}
for scales and axion masses of observational interest. We set $m/H_* = 10$ which is both computationally fast and reaches subpercent accuracy at the half-amplitude scale $k_{1/2}$ as we detailed in \S\ref{sec:method}. In practice, we cease the computation when radiation is $0.1\%$ of the energy density of matter rather than at $a=1$, since for all $k$ of interest $k/a m$ is then sufficiently close to zero that $T_{\rm rel}$ no longer evolves. 

We compare our approach to a naive fluid approximation which solves the fluid equations \eqref{eq:ef_bg} and \eqref{eq:ef_pert} at all times while approximating the equation of state and sound speed with the interpolating form $w_{\rm interp}$ \eqref{eq:wax_interpolate} and the field $c_{s\phi}$ \eqref{eq:cs2_field}. This procedure is similar to the one used by Ref.~\cite{Poulin:2018dzj} and implemented in \textsc{AxiCLASS} with the sound speed $c_{s, \textrm{HN}}$ \eqref{eq:cs2_HN}. We allow the naive fluid approach to use the true $a_c$ when evaluating $w_{\rm interp}$ rather than implement an iterative approximation for it.

Our approach is no more computationally expensive than the naive calculation because with our switch to the EFA at $m/H_* = 10$ we bypass the computationally troublesome axion oscillations (see Fig.~\ref{fig:background}). Nonetheless our scheme is significantly more accurate. In Appendix \ref{app:axioncamb}, we show that the solution scheme used by \textsc{AxionCAMB} yields similar results to the naive fluid approximation we focus on here.

We show $T_{\rm rel}$ in Fig.~\ref{fig:Tk} for a range of relevant axion masses. Our approach and the naive fluid approximation agree qualitatively that axions suppress small-scale clustering relative to CDM. In detail, however, the naive fluid approximation slightly but systematically underestimates the suppression scale of the power spectrum.

\begin{figure}[t]
\includegraphics[]{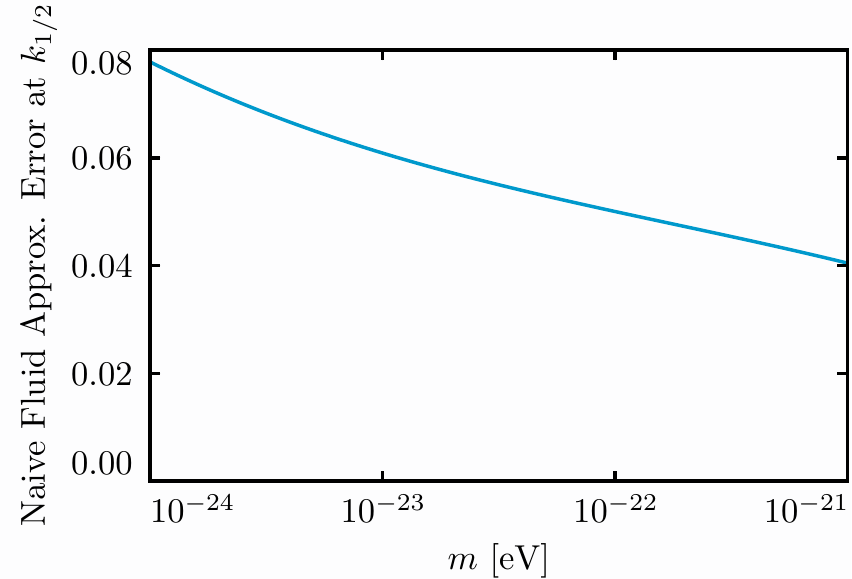}
\caption{The fractional error in the relative axion/CDM transfer function $T_{\rm rel}$ made by the naive fluid approximation at the half amplitude point $k_{1/2}$. The naive fluid approximation makes a systematic error of several percent which increases at low mass. For further discussion see \S\ref{sec:results}.}
\label{fig:fluid_acc}
\end{figure}

In Fig.~\ref{fig:fluid_acc}, we quantify the error in $T_{\rm rel}$ made by the naive fluid approximation at the half-amplitude point $k_{1/2}$, as a function of axion mass. The naive fluid approximation makes a $\sim 4\%$ error for a $10^{-21}\ \eV$ axion and becomes less accurate at lower masses. At $10^{-24}\ \eV$ it makes a nearly $10\%$ error.

While the error made by the naive fluid approximation is significant, it may appear to be smaller than the order unity errors we might have have expected based on extrapolating the scalings we showed in Fig.~\ref{fig:background_fluid_accuracy} and Fig.~\ref{fig:pert_fluid_accuracy} backwards to a switch epoch $m/H \ll 1$.

In fact, for a fixed axion field displacement the naive fluid approximation at the level of the background does make an order unity error in the axion density today, as we would expect from Fig.~\ref{fig:background_fluid_accuracy}. However, since we have normalized the density today to that of dark matter and allowed the initial axion field displacement to float, this error is factored out of our results here. Nonetheless if one wishes to relate the axion initial conditions to the final density, the naive fluid approximation makes an order unity error. 

Likewise at the level of the perturbations, our analysis of evolution errors from the sound speed and matching errors in the presence of Jeans oscillations might seem to imply an even larger discrepancy for both the naive fluid approximation and \textsc{AxionCAMB} (see App.~\ref{app:axioncamb}), which switches at $m/H_*=3$, but these are mitigated by metric sourcing as discussed in \S\ref{subsubsec:matching_pert}.

For convenient comparison to our results and to enable their use in initializing nonlinear simulations, we construct a fitting function to our EFA results which describes the suppression of the axion transfer function relative to CDM before the first node of the transfer function, described by the empirical form\footnote{This improves on the form given in Ref.~\cite{Hu:2000ke}, where in particular the asymptotic suppression was given as $k^{-8}$ instead of $k^{-6}$ due to fitting the differences in form in the intermediate region.}
\begin{equation}
\label{eq:fitting}
T_{\rm rel}(k) \simeq \frac{\sin(x^{n})}{x^n (1 + B x^{6-n})}\,,
\end{equation}
where
\begin{equation}
x \equiv A \frac{k}{k_J },\quad k_J = 9 m_{22}^{1/2}\,,
\end{equation}
with $m_{22} \equiv m / 10^{-22} \ \eV$. The power law index $n = 5/2$ is mass-independent, while $A$ and $B$ run with the mass as
\begin{align*}
A &= 2.22 m_{22}^{1/25- 1/1000\ln(m_{22})}\,, \\ 
B &= 0.16 m_{22}^{-1/20}\,. \numberthis
\end{align*}

In Fig.~\ref{fig:fitting_function}, we show how this fitting function reproduces our EFA results for axions of mass $m=10^{-21} \ \eV$. The fitting function is designed to be accurate at the $10^{-2}$ level relative to CDM and hence fractionally accurate at the percent level only up to approximately the half-amplitude point $k_{1/2}$, with larger fractional errors once the axions are Jeans suppressed. In terms of mass, the fitting functions were constructed using the transfer function for axions of mass $10^{-24} \ \eV \leq m \leq 10^{-21} \ \eV$.

\begin{figure}[t]
	\includegraphics[]{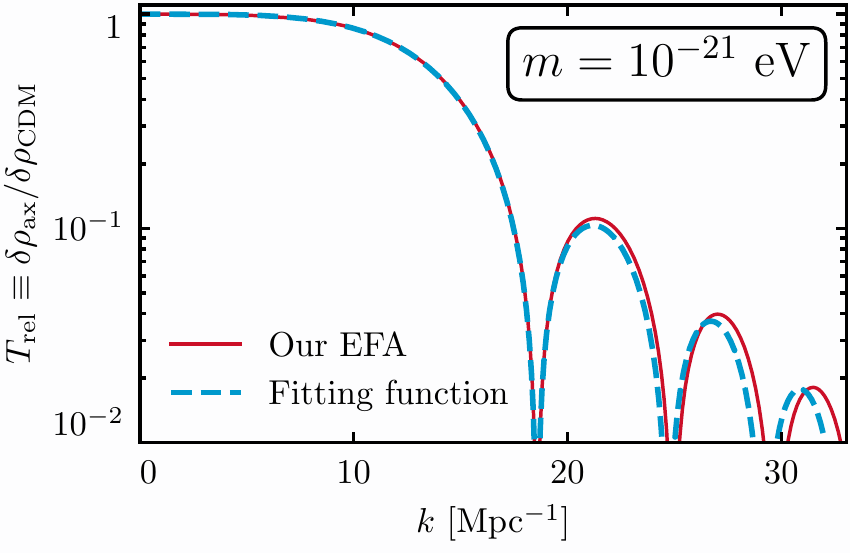}
	\caption{A comparison of our fitting function to our EFA for a $10^{-21} \ \eV$ axion. The fitting function is designed to be fractionally accurate at better than the percent level up to roughly the half amplitude scale $k_{1/2}$ for axions masses between $10^{-24} \ \eV$ and $10^{-21} \ \eV$. Larger fractional errors are made on smaller scales and especially after the first node induced by the Jeans suppression. For further discussion see \S\ref{sec:results}.}
	\label{fig:fitting_function}
\end{figure}

We do not seek a higher level of accuracy in fitting $T_{\rm rel}$ since the omission of baryons and neutrinos already makes an error at this level at $k_{1/2}$, which we have verified using a full Einstein-Boltzmann code by varying them in the naive fluid approximation. 
The even larger effect of the two on the axion transfer function itself can be restored by multiplying $T_{\rm rel}$ with the
full CDM transfer function.
The remaining relative {error} reflects only the simplicity of our code rather than any limitation in our axion solution scheme developed in \S\ref{sec:method}, and can easily be rectified by implementing our scheme in a more complete code such as \textsc{CAMB} or \textsc{CLASS}. 

Because the error made by the naive fluid approximation  at $k\lesssim k_{1/2}$  causes a shift in the suppression scale of the power spectrum, the error can be interpreted as a shift in effective axion mass. In Fig.~\ref{fig:mass_shift}, we show that the error in the naive fluid approximation is comparable to a 3\% shift in the axion mass. Thus for accurate percent level cosmological constraints the naive fluid approximation is not suitable and the approach presented in this work should be preferred to it.

\begin{figure}[t]
	\includegraphics[]{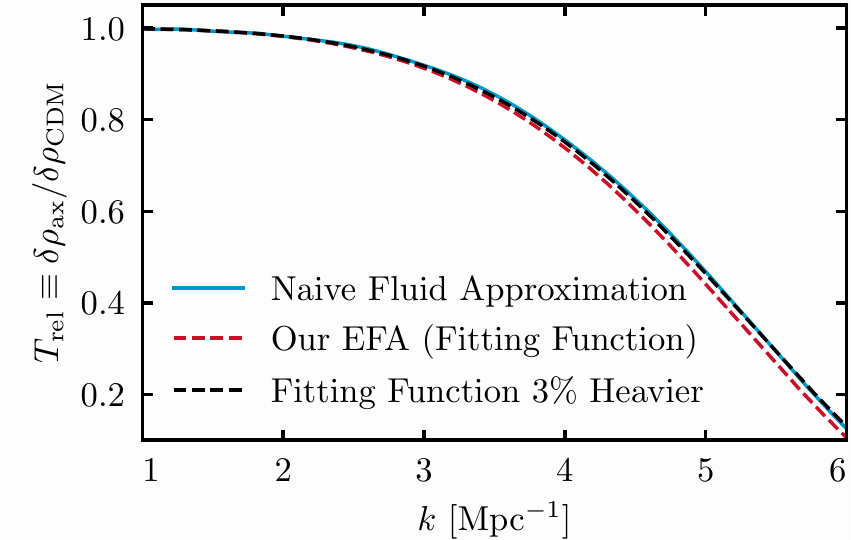}
	\caption{For $m = 10^{-22}\ \eV$, the naive fluid approximation's error at the half-amplitude point $k_{1/2}$ is comparable to a $3\%$ shift in the axion mass. For further discussion see \S\ref{sec:results}.} 
	\label{fig:mass_shift}
	\end{figure}

\section{Conclusion}
\label{sec:conclusion}

	As next-generation cosmological experiments provide precision tests of the ultralight axion dark matter hypothesis and the more general string axiverse idea that ultralight axions should be plentiful, the accuracy of theoretical predictions in these models must increase in parallel.

	These scalar fields induce a new evolution timescale beyond the Hubble rate in the cosmological system associated with their mass. This timescale hierarchy can be challenging to solve, and existing approximation schemes for solutions induce errors which are poorly understood and can be large relative to the precision of cosmological data.

	In this work we have developed a solution procedure for ultralight axions which enables a dramatic reduction in the computation time required to obtain their cosmological observables, as well as an improved theoretical understanding of magnitude and sources of the remaining error. We can achieve subpercent accuracy in observables without having to track even a single full oscillation of the axion field. We can achieve even higher accuracy, at the expense of an increased computation time, by increasing our switch-time parameter $m/H_*$ later and later into the oscillatory regime.

	Our scheme involves matching the axion field to an effective fluid so that they agree at the late times at which observations are made rather than the early times at which the matching is performed. We then approximate the evolution of the effective fluid rather than solving it exactly using an internally calibrated  equation of state and sound speed. It is these two improvements over the existing approaches which enable our massive improvement in accuracy.

	Using our new approach, we were able to quantify the accuracy of existing techniques used in the literature such as the naive approximation which uses fluid equations at all times or the more advanced approach used by \textsc{AxionCAMB}. We showed that these induce errors of several percent or more in 
	density fluctuations relative to an equivalent CDM system today on scales where they are still relatively large, resembling a small mis-scaling of the transfer function with $m_{\rm ax}$.
	
	Ref.~\cite{Cookmeyer:2019rna} already showed that these errors will be significant for CMB-S4 constraints on axions in the mass range $10^{-27}\ \eV \lesssim m_\ax \lesssim 10^{-24}\ \eV$, an issue which can be directly resolved by our approach once it is implemented in a complete Boltzmann solver. For heavier axions, observational tests are in the nonlinear regime and our results provide the linear theory power spectrum before it is processed by nonlinear physics. Since some nonlinear effects can be highly sensitive to the axion mass (see, e.g., Ref.~\cite{Dalal:2022rmp}), rectifying the mis-scaling of the linear theory transfer function with mass using our approach may then be important for self-consistent analysis of nonlinear observations.
	
	Existing approaches can also lead to order unity errors in relating initial axion parameters to final parameters today, and therefore our results can be of use to studies which require proper understanding of, e.g., the axion misalignment angle.

	We provided fitting functions for our results, and our procedure is also
	simple to implement. Relative to the typical approach used in codes like \textsc{AxionCAMB}, our scheme involves just changing the initial conditions used to match to a fluid approximation using Eqs.~\eqref{eq:ansatz}, \eqref{eq:rhoPef}, \eqref{eq:constraints} for the background and Eqs.~\eqref{eq:pert_ansatz}, \eqref{eq:fluidperts}, \eqref{eq:pert_constraints} for the perturbations, and improving the equation of state using \eqref{eq:wax_next} and the sound speed using \eqref{eq:cs2_next}. Doing so in public codes will help unleash the full constraining power of precision cosmology on these fascinating dark sector candidates.

\acknowledgments

	We thank Daniel Grin for fruitful discussions as well as helpful comments on a draft of this work, along with David Zegeye, Evan McDonough, Jose Ezquiaga, Macarena Lagos, and Meng-Xiang Lin for additional insights.

	SP and WH were supported by U.S.\ Dept.\ of Energy contract DE-FG02-13ER41958 and the Simons Foundation.  SP was additionally supported by the Kavli Institute for Cosmological Physics at the University of Chicago through grant NSF PHY-1125897 and an endowment from the Kavli Foundation and its founder Fred Kavli. This work was made possible by the World Premier International Research Center Initiative (WPI), MEXT, Japan. \\

\appendix

\section{AxionCAMB}
\label{app:axioncamb}

\begin{figure}[t]
\includegraphics[]{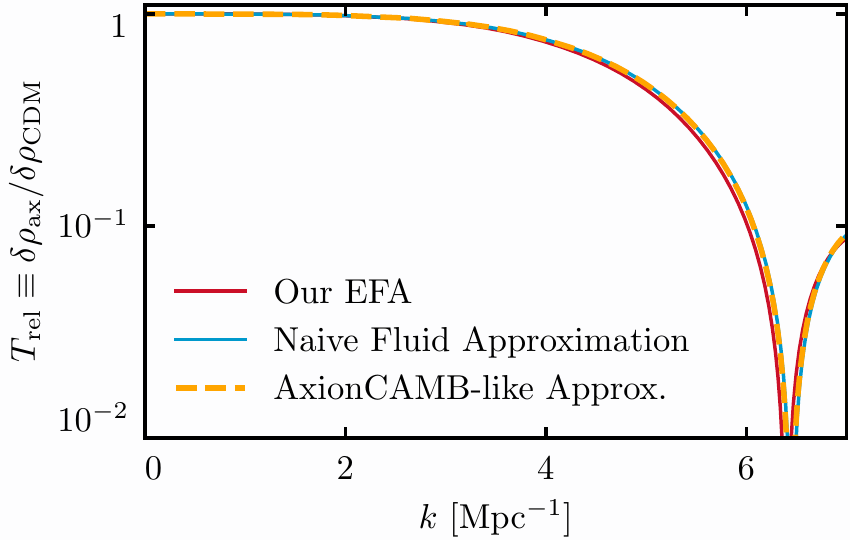}
\caption{The approach used by \textsc{AxionCAMB} yields similar results to the 
naive fluid approximation which we compared to our approach in the main text. Both are discrepant with the more accurate calculation developed in this work. We show here a $10^{-22} \ \eV$ axion with the abundance of dark matter. For further discussion see Appendix~\ref{app:axioncamb}.} 
\label{fig:Tk_camblike}
\end{figure}

In the main text, we compared our EFA to a naive approach which solves the fluid equations at all times with rough approximations for the equation of state and sound speed. In this appendix we show that the approach implemented in the \textsc{AxionCAMB} code yields similar results to the naive fluid approach.

\textsc{AxionCAMB} \cite{Hlozek:2014lca} is a version of the \textsc{CAMB} \cite{Lewis:1999bs} Boltzmann solver which has been modified to include axions. We do not show direct comparisons with \textsc{AxionCAMB} since our code does not include the cosmological effects of baryons and neutrinos on the matter power spectrum. Instead, we reproduce the axion solution scheme that \textsc{AxionCAMB} uses and implement it in our own more simplistic code.

At the background level, \textsc{AxionCAMB} solves exact equations until the onset of axion oscillations at a time $a_{\rm osc}$ defined by
\begin{equation}
m/H\vert_{a_{\rm osc}} \equiv 3 \,.
\end{equation}
After that time, the axion density is evolved in an EFA with $w_\ax=0$.

For the perturbations, at early times $a<a_{\rm osc}$ the fluid equation is solved with $c_s^2 = 1$ and the exact $c_a^2$. Once $a>a_{\rm osc}$, $c_s^2$ takes the Hwang \& Noh EFA form \eqref{eq:cs2_HN}, while $c_a^2$ is set to zero.

We implement this AxionCAMB-like approach in our code and in Fig.~\ref{fig:Tk_camblike} show that it yields results which are very similar to the naive fluid approximation results presented in the main text. In particular it shows the same several-percent level error at $k_{1/2}$.

\clearpage

\bibliographystyle{apsrev4-1}
\bibliography{references.bib}

\end{document}